\documentclass[11pt]{article}
\usepackage{jinstpub}

\usepackage{graphicx}
\usepackage[title]{appendix}
\title{Fabrication and Characterization of p-type Inverted Coaxial Point Contact (ICPC) Detectors with a-Ge Dual-Blocking Contacts}

\author[a]{S.~A.~Panamaldeniya, K.~M.~Dong, and D.~M.~Mei}
\affiliation[a]{Department of Physics, University of South Dakota, Vermillion, USA, 57069}

\emailAdd{dongming.mei@usd.edu}

\abstract{We report the fabrication and characterization of two p-type inverted coaxial point contact (ICPC) high-purity germanium (HPGe) detectors, SAP16 and SAP17, produced from USD-grown crystals with net impurity concentrations of $\sim 3\times10^{10}\,\mathrm{cm^{-3}}$. Both devices employ \emph{thin} amorphous-germanium (a-Ge) dual-blocking contacts, implemented here for the first time on ICPC detectors, to provide bipolar charge blocking while limiting dead-layer thickness. Electrical tests at 76~K demonstrate stable operation with picoampere-level leakage currents and sub-pF capacitance: SAP17 reached $\sim 4.62$~pA at the maximum tested bias (500~V) and operated stably at 400~V with $C\simeq 0.503$~pF. \emph{Meanwhile,} SAP16 achieved superior spectroscopic performance, with energy resolutions of 2.42\% at 59.5~keV and 0.36\% at 662~keV. Gamma-ray spectroscopy with $^{241}$Am and $^{137}$Cs shows that modest geometric differences lead to measurable changes in depletion behavior and charge-collection uniformity, consistent with electrostatic modeling. Angular-response measurements further reveal pronounced directional sensitivity at 59.5~keV, whereas the 662~keV response is essentially isotropic over the measured range. These results validate thin a-Ge dual-blocking contacts for ICPC HPGe detectors and highlight geometry-driven trade-offs among leakage current, depletion, and energy resolution relevant to low-background and low-threshold applications.}


\begin{document}
\maketitle
\flushbottom

\section{Introduction}
Over the past three decades, planar point contact (PPC), Broad Energy Ge (BEGe), and inverted coaxial point contact (ICPC) high-purity germanium (HPGe) detector geometries have been developed to meet the increasingly stringent requirements of rare-event searches, including neutrinoless double-beta decay and dark-matter experiments \cite{Agnese2018_SCDMS,Mei2006_Muon,RomoLuque2025_LEGEND,Agostini2020_GERDA}. These applications demand detectors that simultaneously provide large active volumes for high detection efficiency and extremely low capacitance ($<1\,\mathrm{pF}$) to suppress electronic noise and enable low-energy sensitivity \cite{HeidelbergMoscow2000,Barbeau2007_ULGe}. Conventional coaxial HPGe detectors support large crystal masses but suffer from high electrode capacitance (typically 30--50~pF) \cite{Brudanin2011_HPGe}, which limits low-energy performance \cite{Knoll2010_RDM}, while PPC detectors achieve sub-pF capacitance through small readout electrodes \cite{Barton2016_ULNGe,Wei2022_PPC} but are typically restricted to sub-kilogram ($\sim1\,\mathrm{kg}$) masses \cite{LEGEND1000_PCDR}. The ICPC geometry, first proposed by Cooper et al.\ \cite{Cooper2011_NovelHPGe}, resolves this trade-off by preserving a true point contact readout while accommodating kilogram-scale ($\sim2$--$3\,\mathrm{kg}$) detector volumes, thereby combining low electronic noise with large active mass and excellent pulse-shape discrimination capability \cite{LEGEND1000_PCDR,Cooper2011_PSA,Domula2018_ICPC_PSD}.

The performance of an HPGe detector depends on several factors, including impurity concentration \cite{Meng2019_aGe,Li2020_GeFiCa}, detector geometry \cite{Li2020_GeFiCa}, and critically the choice of electrical contacts \cite{Amman2018_aGe}. Most ICPC detectors used in current rare-event programs employ a lithium-diffused $n^{+}$ outer contact together with a boron-implanted $p^{+}$ contact \cite{Bertoldo2021_nplus,Bhatt2014_GeJunction}. The lithium-diffused $n^{+}$ contact typically introduces a $\sim1$~mm dead layer and an additional $\sim1$~mm transition layer. While the dead layer \cite{Kaya2022_DeadLayer, Andreotti2014} is beneficial for blocking surface $\alpha$ and $\beta$ particles, it reduces the active mass of costly enriched $^{76}$Ge directly impacting the exposure achievable for neutrinoless double-beta decay searches. The transition layer can also prolong charge-drift times and promote incomplete charge collection, which may degrade energy resolution and pulse-shape discrimination. Moreover, lithium-diffused contacts evolve with time and thermal history due to continued lithium migration \cite{Quang2011_DeadLayer,Wang2015_HPGeGrowth}, and they require high-temperature processing steps that complicate fabrication and limit flexibility when prototyping novel geometries.

Thin amorphous semiconductor contacts, particularly amorphous germanium (a-Ge), offer an attractive alternative because they can provide bipolar blocking (suppressing both electron and hole injection) while avoiding high-temperature diffusion processing \cite{Amman2018_aGe,Vetter2007_GeReview,Meng2019_aGe}. When properly fabricated, a-Ge contacts provide effective surface passivation and stable leakage suppression \cite{Amman2007_aGeContacts,Looker2015_Leakage}, and importantly introduce negligible dead and transition layers compared to lithium-diffused $n^{+}$ contacts. This can preserve active enriched material and improve charge collection, offering a pathway to enhanced spectroscopic performance and more uniform electric-field behavior in ICPC detectors. The trade-off is that thin contacts do not intrinsically self-shield against surface $\alpha/\beta$ backgrounds; instead, they enable these events to be measured and discriminated. In experiments that combine HPGe detectors with an active veto such as liquid argon (e.g., LEGEND-200 and LEGEND-1000), thin-contact ICPC detectors may therefore provide improved overall background rejection while also reducing gamma-ray response distortions associated with energy loss in thick $n^{+}$ dead layers.

In this work, we present the fabrication and comprehensive characterization of two p-type ICPC HPGe detectors, SAP16 and SAP17, fabricated from USD-grown HPGe single crystals produced by the Czochralski method \cite{wang2014dislocation} using zone-refined germanium ingots and employing a-Ge dual-blocking contacts \cite{Wei2022_PPC,Wang2015_HPGeGrowth}. Although the two devices share the same ICPC architecture and contact technology, deliberate geometric differences allow us to examine how detector geometry influences depletion behavior, leakage current, and spectroscopic performance in practice. Through a systematic comparison of SAP16 and SAP17, we investigate geometry-dependent trade-offs between electrical stability and energy resolution. We present temperature-dependent electrical measurements, capacitance extraction, gamma-ray spectroscopy using $^{241}$Am and $^{137}$Cs sources, electrostatic field and potential simulations, and angular-response studies, which together provide experimental insight into limitations and design considerations critical for optimizing future ICPC HPGe detectors for low-noise rare-event applications. The remainder of this paper is organized as follows: Section~\ref{sec:DETECTOR DESIGN AND FABRICATION} describes the detector geometry and fabrication process; Section~\ref{sec:Experimental setup and analysis methods} outlines the experimental setup and analysis methods; Sections~\ref{sec:Electrical characterization} and \ref{sec:Gamma-ray spectroscopy performance} present the electrical and spectroscopic characterization results; Section~\ref{sec:Electrostatic simulation and interpretation} discusses electrostatic simulations; Section~\ref{sec:Angular response and relative efficiency} presents angular-response measurements; and Sections~\ref{sec:Discussion} and \ref{sec:Conclusion} provide discussion and conclusions.

\section{Detector design and fabrication}
\label{sec:DETECTOR DESIGN AND FABRICATION}

\subsection{Geometry, fabrication flow, and contact formation}

The SAP16 and SAP17 detectors were fabricated in the p-type inverted coaxial point contact (ICPC) geometry from USD-grown, zone-refined HPGe single crystals. Key geometrical and fabrication parameters are summarized in Table~\ref{tab:geometry_parameters}. In the ICPC configuration, the detector body is a right circular cylinder with a shallow coaxial bore machined from the bottom surface and a small point contact centered on the top surface. The outer cylindrical surface together with the bore wall form the opposing electrode that surrounds the active volume, enabling large detector volumes while preserving the low-capacitance point contact readout.

\begin{figure}
    \centering
    \includegraphics[width=1\linewidth]{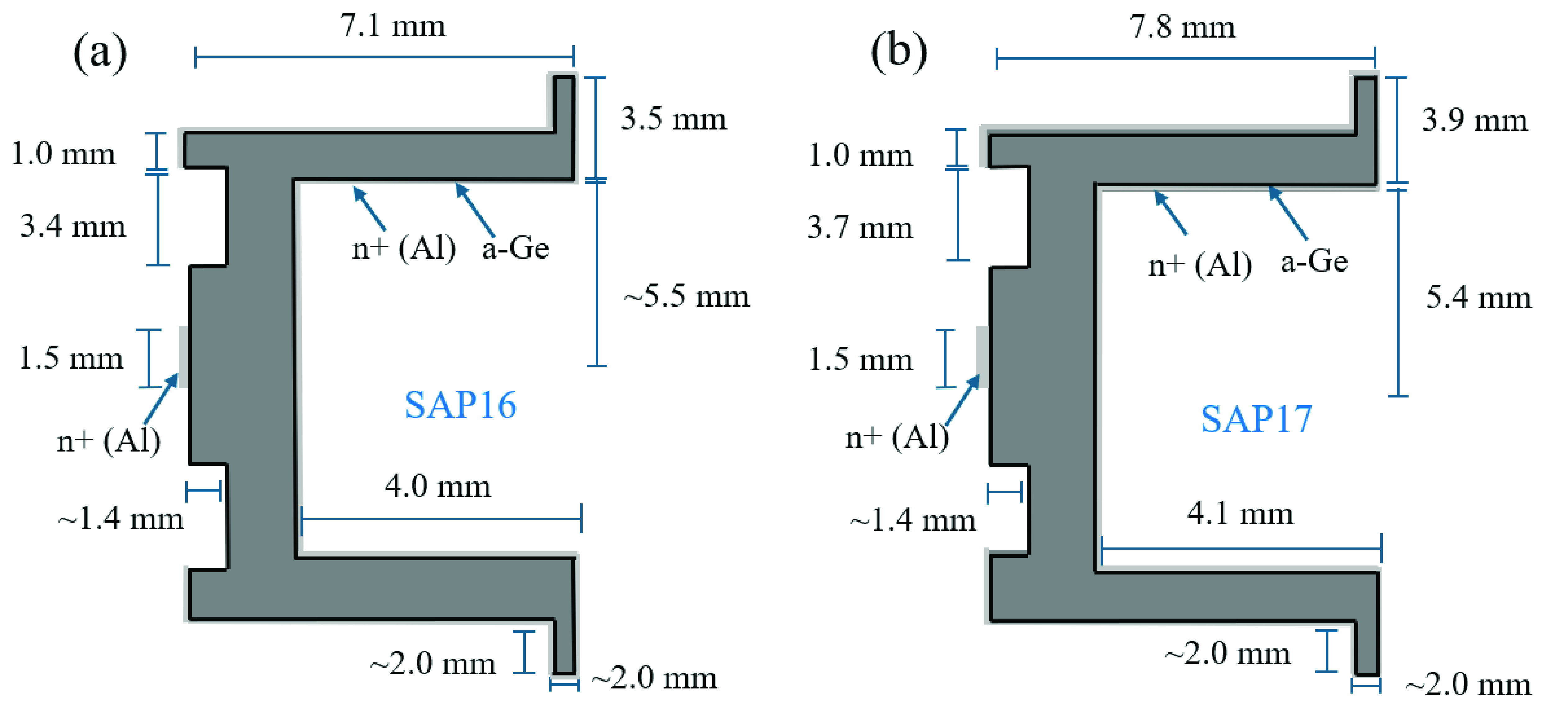}
    \caption{Schematic diagrams of both (a) SAP16 and (b) SAP17 inverted coaxial detectors.}
    \label{Detector diagrams}
\end{figure}

\begin{figure}
    \centering
    \includegraphics[width=1\linewidth]{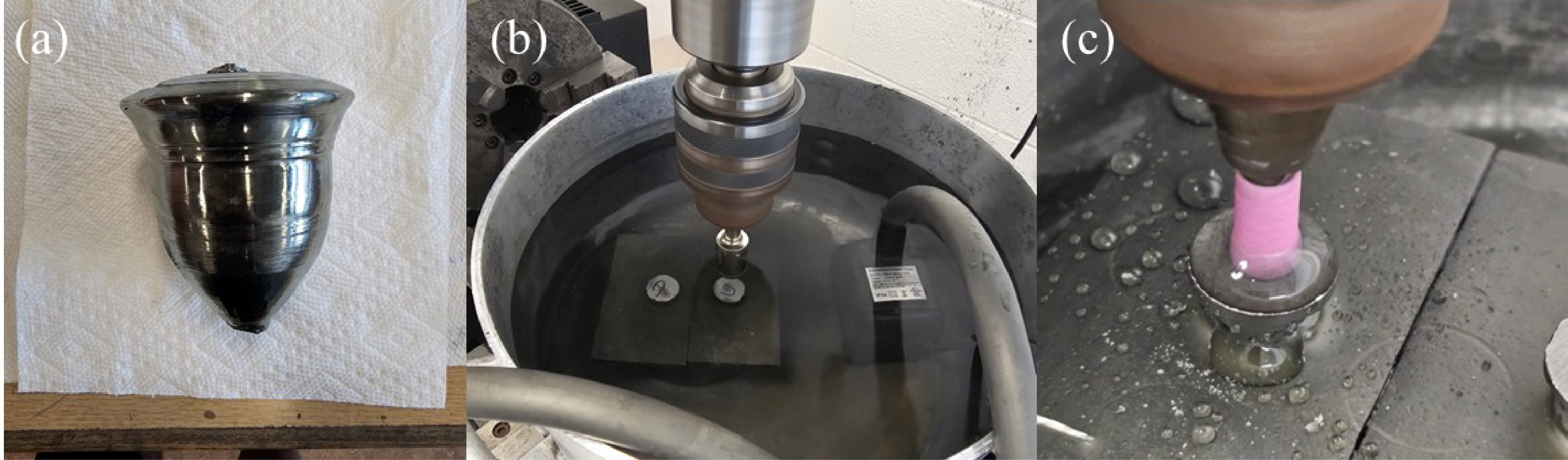}
    \caption{Fabrication steps of ICPC detectors:
    (a) as-grown HPGe crystal from Czochralski method,
    (b) machining of the groove and wing structures on the top surface using a lathe machine, and
    (c) cold drilling of the central coaxial bore.}
    \label{bore and groove cutting}
\end{figure}

Two geometric features were incorporated to improve manufacturability and electrical isolation. First, a \emph{winged} structure was added at the bottom surface to facilitate mechanical handling, masking, and reproducible placement during sputter deposition; to our knowledge, this winged handling geometry is implemented here for the first time in ICPC detector fabrication. Second, a circumferential groove was machined near the top contact region to increase the surface leakage path length and electrically isolate the point contact from the surrounding surface, mitigating surface-current contributions and improving bias stability. After machining and surface preparation, all exposed Ge surfaces were coated with amorphous Ge (a-Ge) for passivation and bipolar charge blocking, while metallic contacts were defined only at the point contact and outer-electrode regions. Schematic representations of SAP16 and SAP17 are shown in Figure~\ref{Detector diagrams}, and photographs illustrating key fabrication stages are provided in Figure~\ref{bore and groove cutting}.

\begin{table*}[htp]
\centering
\caption{Geometrical and fabrication parameters of the ICPC HPGe detectors}
\label{tab:geometry_parameters}
\begin{tabular}{lcc}
\hline
\textbf{Parameter} & \textbf{SAP16} & \textbf{SAP17} \\
\hline
Crystal material & p-type HPGe & p-type HPGe \\
Detector body diameter (mm) & 14.0 & 14.4 \\
Detector height (mm) & 7.1 & 7.8 \\
Coaxial bore diameter (mm) & 11.0 & 10.8 \\
Coaxial bore depth (mm) & 4.0 & 4.1 \\
Point contact diameter (mm) & 1.5 & 1.5 \\
Upper wing thickness (mm) & 1.0 & 1.0 \\
Lower wing thickness (mm) & 2.0 & 2.0 \\
Circumferential groove depth (mm) & 1.5 & 1.5 \\
a-Ge coated surfaces & All exposed Ge surfaces & All exposed Ge surfaces \\
Metallized surfaces & Point contact and outer contact & Point contact and outer contact \\
\hline
\end{tabular}
\end{table*}

\subsection{Fabrication procedure} 
\begin{figure}
    \centering
    \includegraphics[width=1\linewidth]{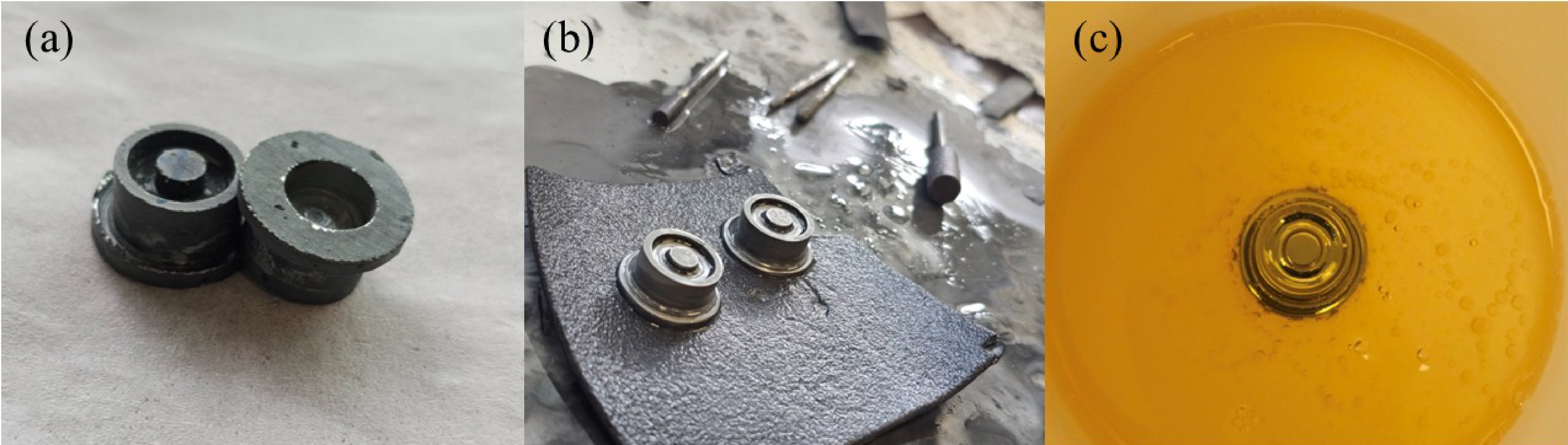}
    \caption{Surface finishing steps of the ICPC HPGe detectors:
(a) detector after groove and bore cutting,
(b) detector after mechanical polishing of all surfaces, and
(c) detector during chemical etching.}
    \label{Surface finishing}
\end{figure}

\begin{figure}
    \centering
    \includegraphics[width=1\linewidth]{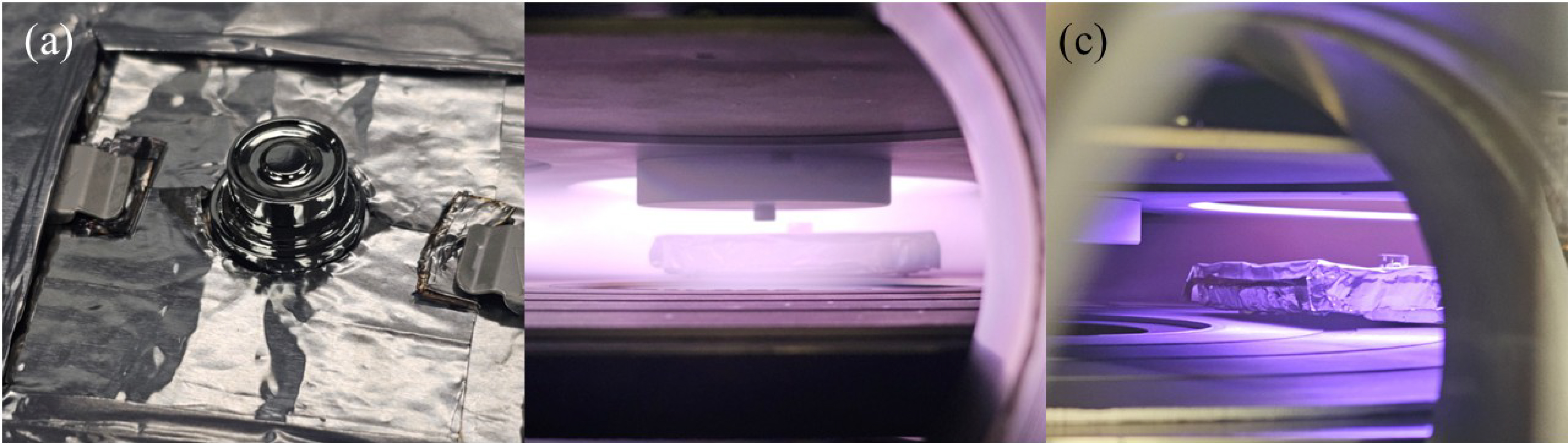}
    \caption{Fabrication steps associated with contact deposition for the ICPC HPGe detectors:
(a) detector loaded into the sputtering chamber after surface treatment,
(b) amorphous germanium (a-Ge) deposition for surface passivation, and
(c) aluminum contact deposition by DC magnetron sputtering under an argon plasma.}
    \label{Deposition}
\end{figure}

\begin{figure}
    \centering
    \includegraphics[width=1\linewidth]{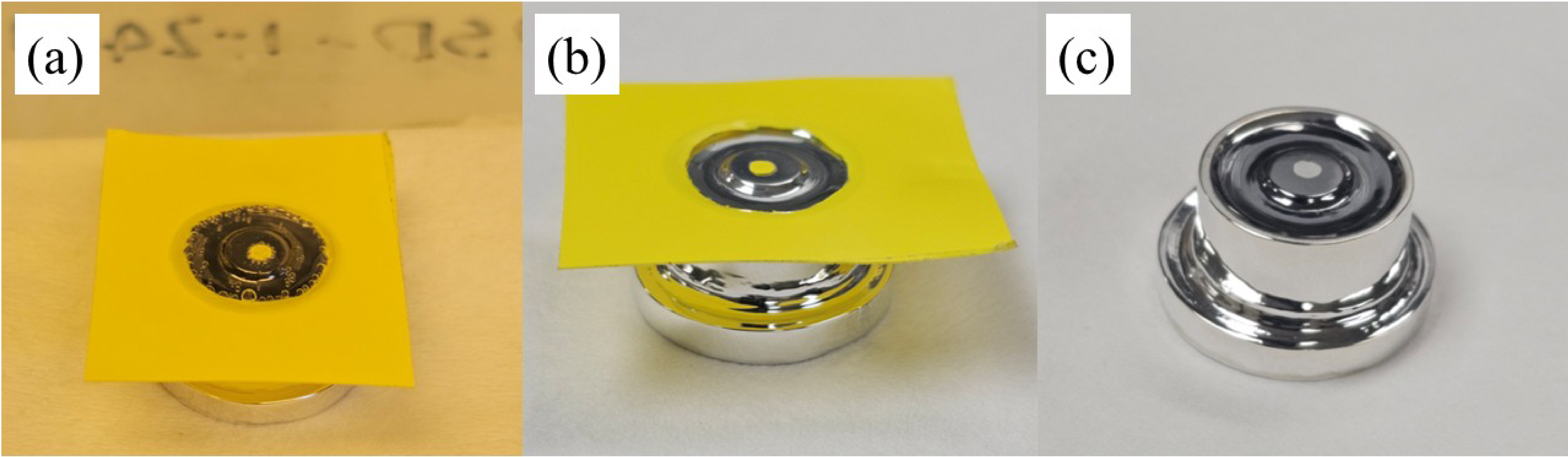}
    \caption{Chemical etching steps of the ICPC HPGe detector:
  (a) masked detector before etching,
  (b) surface condition immediately after HF:HNO$_3$ etching,
  and (c) detector after cleaning and drying.}
    \label{Etching}
\end{figure}

Electrical contacts and surface passivation were fabricated through a controlled sequence of (i) mechanical surface preparation, (ii) chemical etching, (iii) amorphous-Ge deposition, and (iv) metal contact deposition and definition. Following mechanical shaping, surface damage introduced during machining was removed by sequential polishing and chemical etching. The top and bottom faces were lapped using 17.5~$\mu$m followed by 9.5~$\mu$m Al$_2$O$_3$ slurry, while the cylindrical sidewall and bore surfaces were polished using successive SiC abrasive papers with grit sizes of 400, 800, 1250, and 2500. After polishing, the detectors were etched in an HF:HNO$_3$ (1:4) solution (hydrofluoric acid: Thermo Scientific ACS Reagent, 48--51\%; nitric acid: Fisher Chemical A200C-212, ACS Plus) for 6~minutes with constant agitation to remove residual subsurface damage and surface contaminants. Immediately after etching, the detectors were thoroughly rinsed in deionized (DI) water and dried under nitrogen flow. The surface-finishing procedure is illustrated in Figures~\ref{bore and groove cutting}, \ref{Surface finishing}, \ref{Deposition}, \ref{Etching}.

Amorphous germanium (a-Ge) films were deposited on all exposed crystalline-Ge surfaces using a Perkin-Elmer 2400 sputtering system (Figure~\ref{Detector full setup} (b)) under a 7\% H$_2$/Ar gas mixture. Depositions were performed at room temperature (no intentional substrate heating) with a base pressure of $3 \times 10^{-6}$~Torr and an RF power of 100~W. We note that modest substrate temperature variation may occur during sputtering due to plasma heating. The nominal a-Ge thickness was $\sim 600$~nm, providing surface passivation and dual blocking of electron and hole injection.

Aluminum contacts were deposited by DC magnetron sputtering in pure argon at a chamber pressure of $3 \times 10^{-6}$~Torr for 6~minutes using a DC power of 300~W, yielding a nominal thickness of $\sim 120$~nm. Aluminum was selected to provide a low-resistance electrical interface compatible with the underlying a-Ge contact and robust coupling during cryogenic operation. The contact-deposition stages are shown in Figure~\ref{Deposition}.

The point contact was defined by removing the guard-ring region surrounding the top electrode using a localized 1\% HF etch (Hydrofluoric Acid: Thermo Scientific ACS Reagent 48-51\%) applied through a Kapton masking template. This procedure produced a well-defined point contact electrode with a diameter of approximately 1.5~mm that was electrically isolated from the surrounding passivated surface, as shown in Figure~\ref{Etching}c. Electrical continuity of the contacts and isolation of the point contact were verified using resistance measurements prior to cryogenic testing.

No post-deposition annealing or thermal baking was performed after a-Ge or aluminum deposition. Because the a-Ge films can be mechanically fragile, detectors were handled only with non-metallic tools and minimal physical contact; during fabrication they were touched exclusively at the machined wing region to avoid damaging the passivated and metalized surfaces.

\section{Experimental setup and analysis methods}
\label{sec:Experimental setup and analysis methods}

\subsection{Cryogenic environment and electronics chain}
All measurements were performed under cryogenic conditions in a temperature-variable cryostat cooled with liquid nitrogen (LN$_2$). Unless otherwise stated, detectors were operated at a stabilized temperature of 76~K. The temperature was monitored and regulated using a Lake Shore 335 temperature controller to ensure reproducible electrical and spectroscopic conditions.

A positive high-voltage bias was applied to the bottom (outer) contact of the p-type ICPC detectors, while the point contact was held at ground potential and used for signal readout. Under this biasing convention, the depletion region grows from the outer contact toward the point contact, and the induced charge signal is collected at the point contact electrode. This configuration is consistent with standard operating practice for p-type point contact detectors and provides a stable polarity for suppressing charge injection at the contacts.

\begin{figure}
    \centering
    \includegraphics[width=1\linewidth]{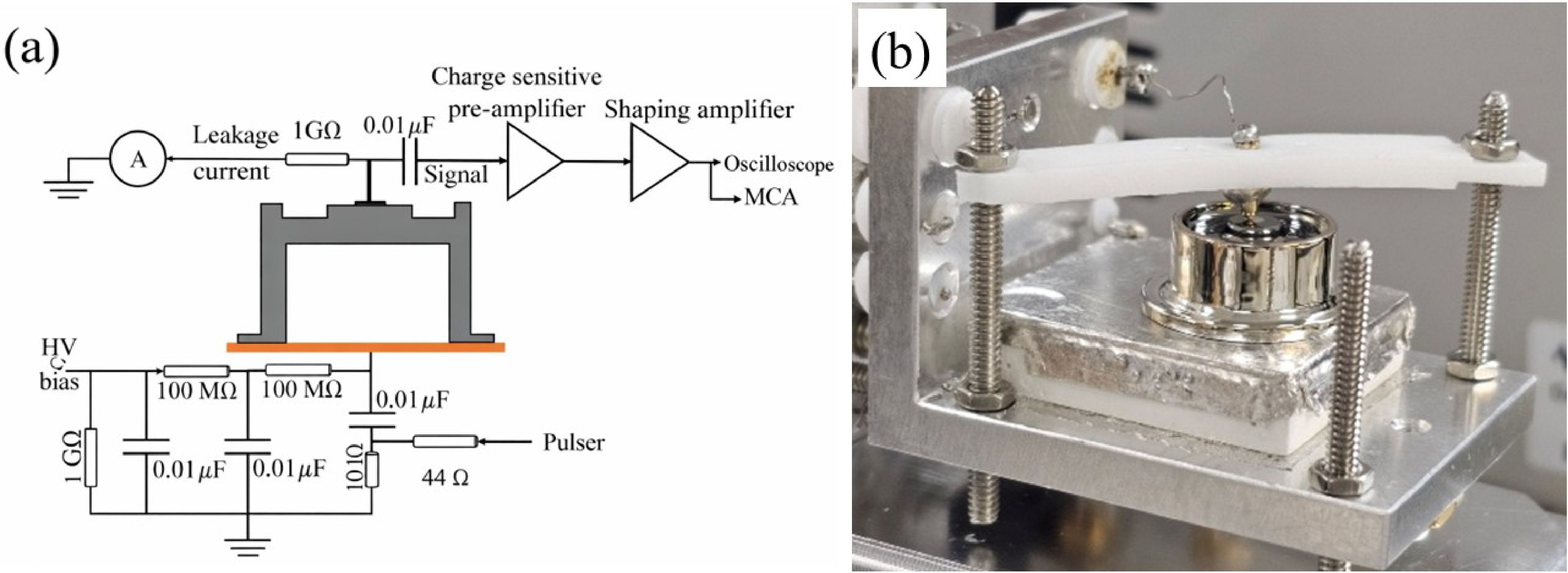}
    \caption{(a) Schematic of the electronics chain used for electrical and spectroscopic measurements and (b) photograph of the detector mounted inside the cryostat}
    \label{electronic setup}
\end{figure}

Leakage currents were measured using a Keithley 6482 dual-channel picoammeter. Gamma-ray spectra were acquired with an ORTEC 927 multichannel analyzer (MCA). Unless otherwise noted, a shaping time of 0.5~$\mu$s was used for spectroscopy to minimize electronic noise while maintaining stable pulse processing. A schematic of the electronics chain and a photograph of a detector mounted inside the cryostat are shown in Figure~\ref{electronic setup}.

\subsection{Capacitance extraction method}
\label{subsec:capacitance_extraction}

The detector capacitance was determined using a charge-injection technique based on a calibrated pulser signal. A pulser with known equivalent energy was injected at the preamplifier input, producing a corresponding voltage step at the preamplifier output. The pulse amplitude was measured at the oscilloscope input using a Keysight InfiniiVision DSOX3034A digital oscilloscope. The detector capacitance was calculated as
\begin{equation}
C = \frac{Q}{V},
\label{eq:capacitance}
\end{equation}
where $Q$ is the injected charge and $V$ is the measured output voltage step amplitude.

The injected charge was obtained from
\begin{equation}
Q = Ne = \frac{E}{w}\,e,
\label{eq:charge}
\end{equation}
where $N$ is the number of electron--hole pairs generated by the pulser, $e$ is the elementary charge ($1.6 \times 10^{-19}$~C), $E$ is the pulser energy expressed in electron volts, and $w$ is the mean energy required to produce one electron--hole pair in Ge at 77~K (2.96~eV) \cite{Looker2014_Thesis}.

The pulser-energy calibration was established using known gamma-ray peaks (662~keV from $^{137}$Cs and 59.5~keV from $^{241}$Am). The dominant sources of uncertainty arise from the pulser-energy calibration and oscilloscope voltage resolution, leading to an estimated relative uncertainty of approximately 6--8\% in the extracted capacitance values, as summarized in Table~\ref{tab:capacitance_comparison}.

\subsection{Energy resolution decomposition}
\label{Energy resolution decomposition}

The total energy resolution of an HPGe detector arises from statistically independent contributions associated with charge-generation statistics, charge-collection variations, and electronic noise. Following the standard treatments of Knoll \cite{Knoll2010_RDM} and Amman \cite{Amman2018_aGe}, these components add in quadrature \cite{Panth2020_CryoGe}:
\begin{equation}
\Delta E_{\mathrm{total}}^{2}
=
\Delta E_{\mathrm{s}}^{2}
+
\Delta E_{\mathrm{c}}^{2}
+
\Delta E_{\mathrm{n}}^{2},
\label{eq:energy_resolution}
\end{equation}
where $\Delta E_{\mathrm{s}}$ is the Fano-limited statistical term, $\Delta E_{\mathrm{c}}$ represents broadening due to incomplete charge collection and/or electric-field non-uniformity, and $\Delta E_{\mathrm{n}}$ is the electronic-noise contribution, which depends primarily on detector capacitance, leakage current, and shaping time.

For point contact and inverted coaxial point contact (ICPC) geometries, the capacitance is extremely small; therefore, $\Delta E_{\mathrm{n}}$ often dominates the measured energy resolution at low energies, while the statistical and charge-collection terms become comparatively smaller.

In this work, the electronic-noise component is determined from the pulser full width at half maximum (FWHM), $\Delta E_{\mathrm{pulser}}$, which provides a direct measure of the front-end noise under the selected shaping conditions. The total detector resolution is obtained from the measured gamma-ray peak FWHM, $\Delta E_{\gamma}$, which represents operational performance under irradiation. To isolate intrinsic detector broadening excluding electronic noise, the pulser contribution is removed via standard quadrature subtraction:
\begin{equation}
\Delta E_{\mathrm{int}}
=
\sqrt{
\left(\Delta E_{\gamma}\right)^{2}
-
\left(\Delta E_{\mathrm{pulser}}\right)^{2}
},
\label{eq:intrinsic_resolution}
\end{equation}
where $\Delta E_{\mathrm{int}}$ includes contributions from Fano statistics, charge-collection variations, and geometry-dependent effects \cite{Knoll2010_RDM}. This intrinsic term is not identical to the purely Fano-limited statistical broadening discussed in \cite{Amman2018_aGe}; rather, it represents the combined non-electronic broadening that remains after removing the electronic-noise contribution.

For reporting, the overall energy resolution is expressed as the relative resolution
\begin{equation}
R(E) = \frac{\Delta E_{\gamma}}{E} \times 100\%,
\label{eq:energy_resolution_ratio}
\end{equation}
where $E$ is the gamma-ray energy. The intrinsic broadening $\Delta E_{\mathrm{int}}$ is used to assess crystal quality, charge-collection uniformity, and the effectiveness of the a-Ge dual-blocking contact technology \cite{Looker2014_Thesis,Looker2015_InterElectrode}.

\section{Electrical characterization}
\label{sec:Electrical characterization}

\subsection{I--V characteristics (leakage current versus bias voltage)}
\label{subsec:I-V characteristics (leakage current versus bias voltage)}

Leakage-current characteristics of the SAP16 and SAP17 ICPC detectors were measured as a function of applied bias voltage at a stabilized temperature of 76~K. At each bias step, the leakage current was recorded after an approximately 1~minute stabilization period; we note that longer waiting times can yield marginally lower currents as surface and interface charge distributions further equilibrate. For all measurements, a positive bias (+HV) was applied to the outer (bottom) contact, while the point contact was held at ground and used for readout.

\begin{figure}
    \centering
    \includegraphics[width=1\linewidth]{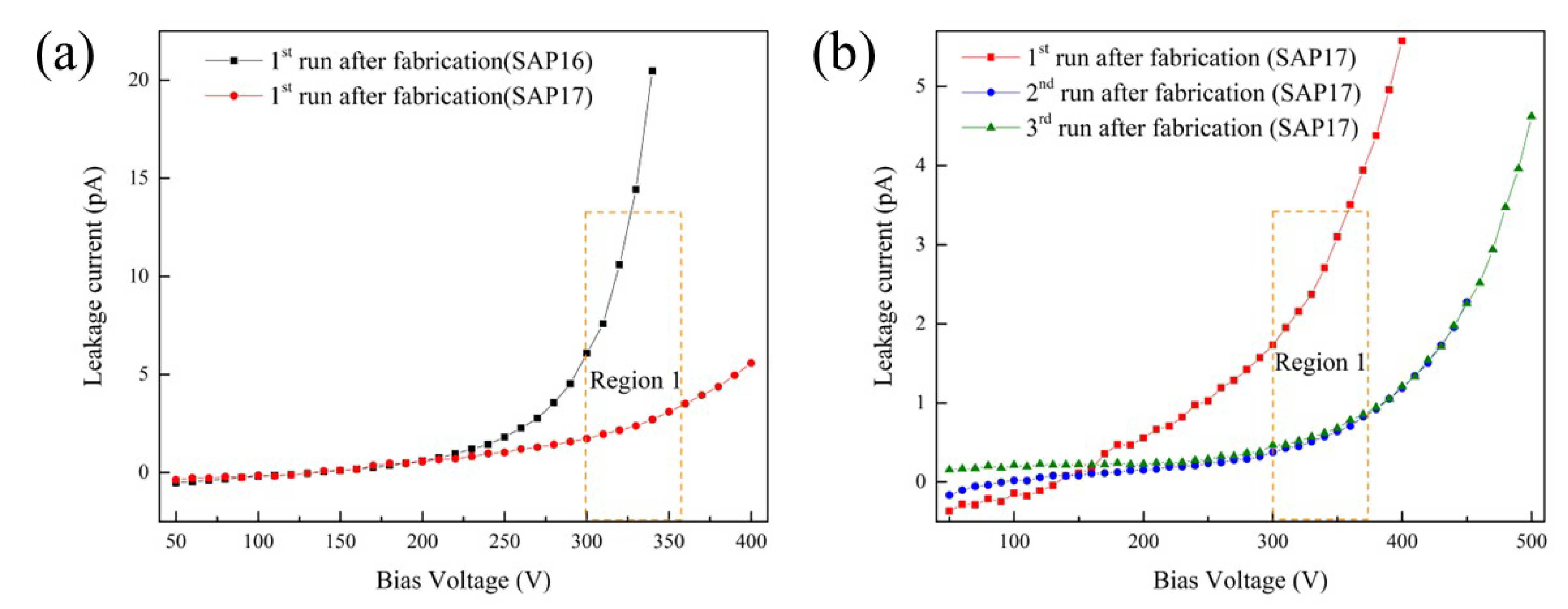}
    \caption{(a) the leakage current as a function of bias voltage for both detectors (SAP16 and SAP17) measured during their first run after fabrication, and (b) the leakage current variation of detector SAP17 measured over three successive runs after fabrication. Region 1 highlights the sudden increase in leakage current, which is attributed to enhanced electron injection into the detector at the point contact after full depletion is reached.}
    \label{Leakage Current of detectors}
\end{figure}

Figure~\ref{Leakage Current of detectors}(a) presents the initial leakage current-voltage (I--V) characteristics of both detectors measured during their first bias sweeps after fabrication. SAP17 exhibits consistently lower leakage current across the full bias range compared to SAP16, while both detectors show the expected increase in leakage current with increasing bias voltage. At their respective operating biases, leakage currents on the order of a few picoamperes were achieved, demonstrating effective charge blocking by the a-Ge contacts.

Based on its lower leakage current, SAP17 was selected for additional stability studies. To assess reproducibility and conditioning behavior, repeated I--V scans were performed on SAP17 over multiple bias cycles, as shown in Figure~\ref{Leakage Current of detectors}(b). The first sweep after fabrication (red symbols) exhibits higher leakage current, which we attribute to a non-equilibrium distribution of trapped charge in the a-Ge contact and passivated surface layers. During the initial one to two bias cycles, the leakage current can show irregular behavior, including small fluctuations and occasionally negative values at low voltages (50-150~V), consistent with incomplete charge equilibration and metastable trapping effects near the contact/interface region. With continued bias cycling, the I--V characteristics converge to lower and highly reproducible leakage-current levels. The close overlap of later scans indicates that the detector reaches a stable operating condition after initial electrical conditioning. We report this behavior as an empirical observation without assigning a unique microscopic mechanism.

Leakage-current measurements were performed under the nominal operating polarity (+HV on the outer contact, point contact at ground), which yielded stable picoampere-level leakage currents. When the bias polarity was reversed, stable operation could not be achieved due to a rapid increase in leakage current, indicating that charge injection is effectively suppressed only under the intended operating polarity for point contact detectors \cite{Wei2022_PPC,Barton2016_ULNGe}. This strong polarity dependence is consistent with electrostatic modeling discussed in Section~\ref{sec:Polarity dependence (Negative bias) and operational choice}.

%Figure~\ref{Temperature data} (a) shows the temperature dependence of the SAP17 leakage current measured between 76~K and 90~K. The leakage current increases monotonically with temperature, consistent with thermally activated carrier transport and enhanced surface leakage. This strong temperature dependence underscores the importance of cryogenic operation for achieving low-noise performance in ICPC detectors.

\begin{figure}
    \centering
    \includegraphics[width=1\linewidth]{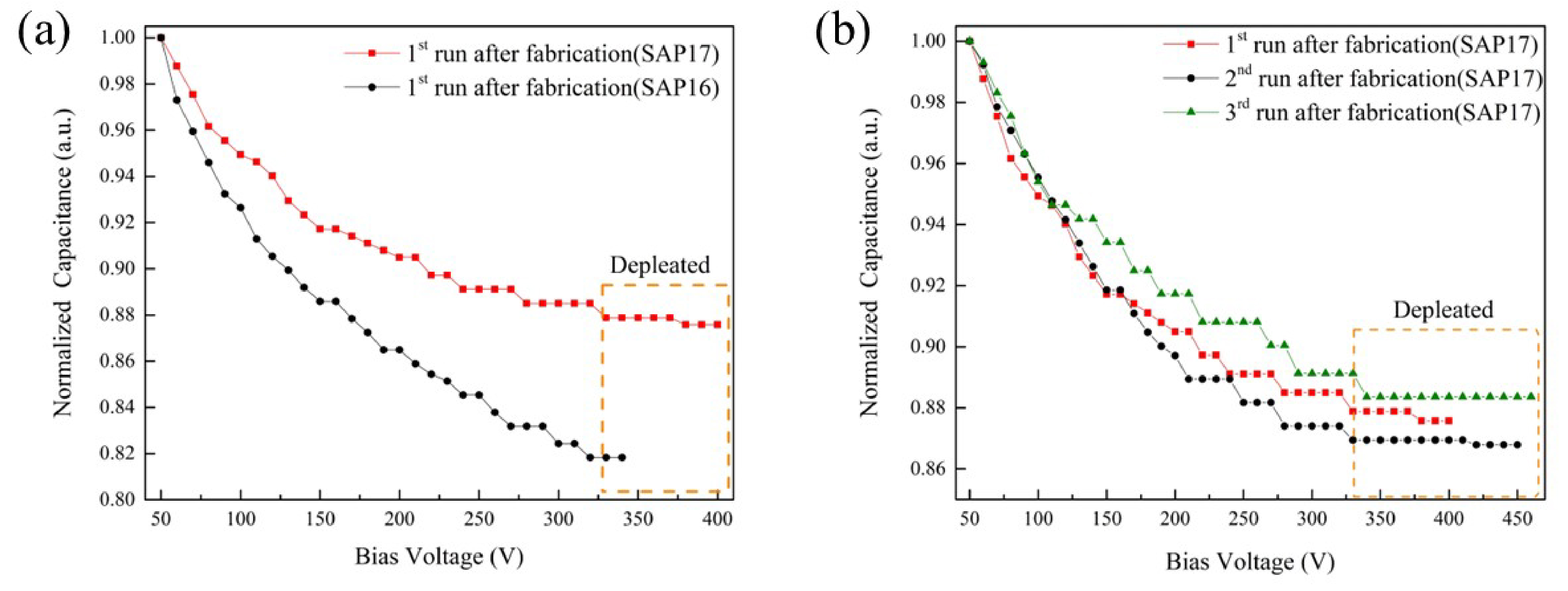}
    \caption{(a) Normalized capacitance as a function of bias voltage for both detectors (SAP16 and SAP17) measured during their first run after fabrication. (b) Normalized capacitance variation of detector SAP17 measured over three successive runs after fabrication. The dashed region indicates the voltage range where the detector is fully depleted.}
    \label{Normalized capacitanceof detectors}
\end{figure}

\subsection{C--V characteristics and depletion assessment}
\label{subsec:C-V characteristics and depletion assessment}

Detector capacitance was measured as a function of applied bias voltage at 76~K
using the charge-injection method described in
Section~\ref{subsec:capacitance_extraction}. Figure~\ref{Normalized capacitanceof detectors}(a)
shows the C--V characteristics of SAP16 and SAP17. For a fully depleted detector
with fixed electrode geometry, the capacitance approaches an approximately
constant value once the depletion region spans the active volume
\cite{Wei2022_PPC}. In these ICPC devices, the normalized capacitance exhibits a
clear change in slope at $V \simeq 330$~V, followed by an approach to a
near-constant value at higher bias.

To assess repeatability and to separate depletion behavior from possible
post-fabrication stabilization effects, SAP17 was re-measured over three
successive C--V runs, as shown in Figure~\ref{Normalized capacitanceof detectors}(b).
In the third run, the capacitance becomes flat (within measurement uncertainty)
for $V \gtrsim 330$~V, consistent with operation at or very near full depletion
under our measurement conditions. The second run shows the onset of a plateau,
while the first run retains a small residual decrease at the highest voltages.
Rather than interpreting the first-run residual slope as definitive evidence of
incomplete depletion, we treat it conservatively as a combination of (i) slow
stabilization of the a-Ge contact/surface charge configuration following
fabrication and initial biasing and (ii) the intrinsic difficulty of defining a
sharp ``full-depletion voltage'' in non-planar ICPC geometries where weak-field
regions can persist near geometric transitions. Accordingly, throughout this work
we use the plateau voltage range ($V \gtrsim 330$~V after stabilization) as an
\emph{operational} indicator that the detector is effectively depleted for
spectroscopy, and we avoid over-interpreting small residual C--V slopes at the
few-percent level.

Electrostatic simulations were performed to further contextualize the C--V
results by examining the spatial distribution of electric potential and field in
the ICPC geometry (Figs.~\ref{Electrostatic} and \ref{potential field},
Section~\ref{sec:Electrostatic simulation and interpretation}). The simulations
predict that weak-field or marginally depleted pockets can remain near the bore
ledge and wing-transition regions at biases comparable to those used in the C--V
measurements. The extent of these pockets is sensitive to the assumed net impurity
concentration and its spatial nonuniformity: the simulations employ a single
uniform impurity value, whereas the effective impurity concentration may vary
with position due to crystal-growth conditions. Taken together, the stabilized
plateau behavior in the measured C--V curves and the simulations support the
interpretation that the detectors operate at or near full depletion in the bulk
at the chosen operating biases, while localized weak-field regions associated
with geometric features may persist and can influence charge collection for some
event topologies.

Capacitance was also simulated using the same Julia-based framework
(\texttt{SolidStateDetectors.jl}) \cite{Abt2021_SSD}. The simulated capacitances
at the operating biases are compared with the extracted values in
Table~\ref{tab:capacitance_comparison}. Agreement is within $\sim$7--8\% for both
detectors, consistent with the expected uncertainty of the pulser-based
capacitance extraction and residual systematic differences between the modeled
and as-built detector/electrode configurations.

\begin{table*}[t]
\centering
\caption{Comparison between experimental and simulated detector capacitances}
\label{tab:capacitance_comparison}
\begin{tabular}{lccc}
\hline
\textbf{Detector (Bias)} &
\textbf{$C_{\mathrm{experimental}}$ (pF)} &
\textbf{$C_{\mathrm{simulated}}$ (pF)} &
\textbf{Difference (\%)} \\
\hline
SAP16 (340~V) & 0.528 & 0.569 & 7.77 \\
SAP17 (400~V) & 0.503 & 0.537 & 6.76 \\
\hline
\end{tabular}
\end{table*}

\section{Gamma-ray spectroscopy performance}
\label{sec:Gamma-ray spectroscopy performance}

\subsection{Representative energy spectra}
Gamma-ray spectroscopic performance was evaluated using standard $^{241}$Am and $^{137}$Cs calibration sources. For each detector, spectra were acquired at the selected operating bias determined from the electrical characterization (SAP16 at 340~V and SAP17 at 400~V), using an identical shaping time of 0.5~$\mu$s and the same operating temperature of 76~K. These operating conditions were chosen to balance low leakage current, stable bias behavior, and low electronic noise.

\begin{figure}
    \centering
    \includegraphics[width=1\linewidth]{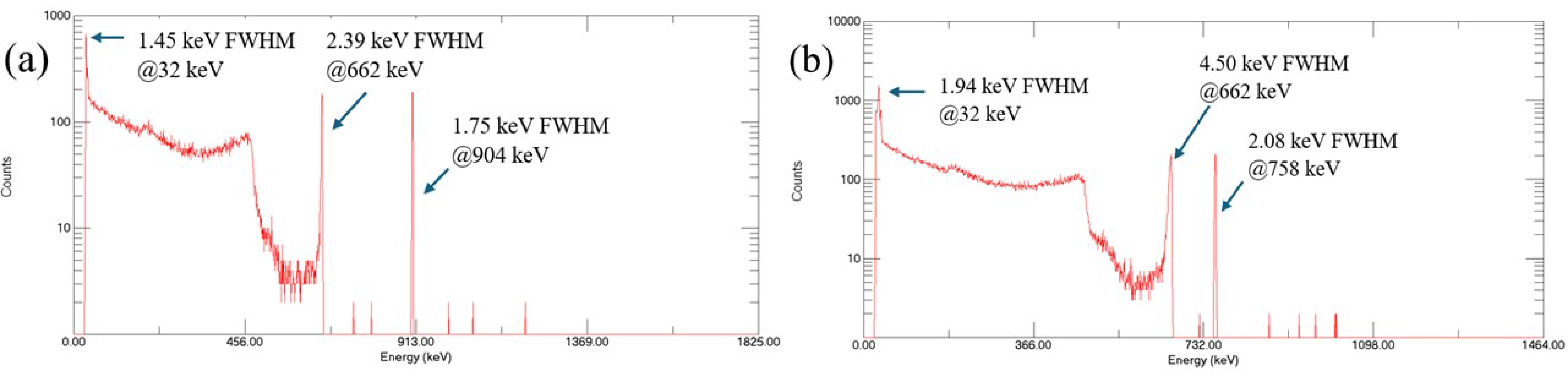}
    \caption{Energy spectra of $^{137}$Cs measured with (a) SAP16 and (b) SAP17 detectors.}
    \label{Cs137}
\end{figure}

\begin{figure}
    \centering
    \includegraphics[width=1\linewidth]{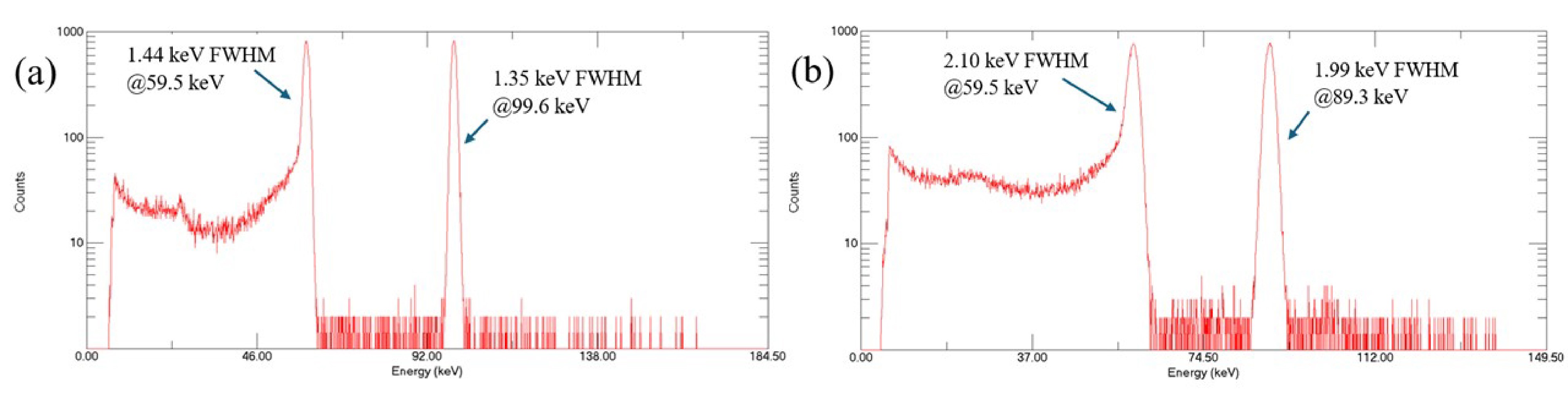}
    \caption{Energy spectra of $^{241}$Am measured with (a) SAP16 and (b) SAP17 detectors.}
    \label{Am241}
\end{figure}

Figure~\ref{Cs137} shows representative $^{137}$Cs spectra obtained with SAP16 and SAP17, and Figure~\ref{Am241} presents the corresponding $^{241}$Am spectra. Both detectors exhibit well-defined photopeaks at 662~keV ($^{137}$Cs) and 59.5~keV ($^{241}$Am), demonstrating stable spectroscopic operation with a-Ge dual-blocking contacts in the ICPC configuration. In addition to the primary $\gamma$-ray photopeaks, both detectors show low-energy features associated with characteristic X-ray interactions and scattering continua. For $^{137}$Cs, the Ba K-shell X-ray peak is visible near 32~keV, consistent with fluorescence following photoelectric absorption in the source or surrounding materials. For $^{241}$Am, the corresponding low-energy X-ray structure is present but not clearly resolved under the present operating and shaping conditions.

Qualitatively, SAP16 exhibits narrower photopeaks and reduced low-energy tailing compared to SAP17 at both energies, indicating improved charge-collection uniformity and reduced non-Gaussian broadening. In contrast, SAP17 shows more pronounced peak broadening and tailing, particularly at 662~keV. This behavior is consistent with a stronger influence of carrier drift-path length and field non-uniformity on the charge-collection process for SAP17, as discussed in the electrostatic simulations (Section~\ref{sec:Electrostatic simulation and interpretation}).

\subsection{Energy resolution summary}
Energy resolution was quantified using the full width at half maximum (FWHM) of the gamma-ray photopeaks. Throughout this work, the measured photopeak width is denoted as $\mathrm{FWHM}_{\gamma}$, the electronic-noise contribution determined from the pulser peak is denoted as $\mathrm{FWHM}_{\mathrm{pulser}}$, and the intrinsic detector contribution obtained by quadrature subtraction is denoted as $\mathrm{FWHM}_{\mathrm{int}}$, as defined in (Section~\ref{Energy resolution decomposition}).

Table~\ref{tab:energy_resolution} summarizes the energy-resolution performance of SAP16 and SAP17 at 59.5~keV and 662~keV, including the applied bias voltage, shaping time, measured photopeak width, pulser width, and extracted intrinsic contribution. SAP16 achieves superior resolution at both energies, with smaller $\mathrm{FWHM}_{\gamma}$ and reduced intrinsic broadening. SAP17 exhibits larger intrinsic broadening, most notably at 662~keV, where the increase in $\mathrm{FWHM}_{\mathrm{int}}$ indicates enhanced sensitivity to charge-collection non-uniformities and field inhomogeneities associated with its geometry and incomplete depletion under the maximum stable bias. At 59.5~keV, both detectors remain largely noise-dominated, while at 662~keV the intrinsic term becomes a more significant fraction of the total width for SAP17, consistent with increased charge-transport path lengths and geometry-dependent collection effects.

\begin{table*}[t]
\centering
\caption{Energy resolution components of the SAP16 and SAP17 detectors at different gamma-ray energies.}
\label{tab:energy_resolution}
\resizebox{\textwidth}{!}{%
\begin{tabular}{lcccccc}
\hline
Detector & Energy (keV) & Bias (V) &
FWHM$_{\mathrm{pulser}}$ (keV) &
FWHM$_{\gamma}$ (keV) &
FWHM$_{\mathrm{int}}$ (keV) &
Resolution (\%) \\
\hline
SAP16 & 59.5 & 340 & 1.35 & 1.44 & 0.50 & 2.42 \\
SAP16 & 662  & 340 & 1.75 & 2.39 & 1.63 & 0.36 \\
SAP17 & 59.5 & 400 & 1.99 & 2.10 & 0.67 & 3.53 \\
SAP17 & 662  & 400 & 2.08 & 4.50 & 3.99 & 0.68 \\
\hline
\end{tabular}}
\end{table*}

\section{Electrostatic simulation and interpretation}
\label{sec:Electrostatic simulation and interpretation}

\subsection{Simulation framework and assumptions}

Electrostatic simulations were performed to evaluate the electric potential and field distributions within the SAP16 and SAP17 ICPC detector geometries and to aid interpretation of the measured electrical and spectroscopic behavior. Three-dimensional detector models were constructed to match the physical dimensions summarized in Table~\ref{tab:geometry_parameters}, including the shallow coaxial bore, point contact location, wing structures, and the circumferential surface groove. The simulations assume a uniformly doped p-type Ge crystal with an effective acceptor concentration consistent with the measured impurity level of approximately $3 \times 10^{10}\,\mathrm{cm^{-3}}$. All exposed Ge surfaces were treated as electrically passivated, and boundary conditions were applied to represent the biased outer electrode and grounded point contact. Simulations were performed at bias voltages corresponding to the operating conditions used experimentally (340~V for SAP16 and 400~V for SAP17).

For interpretation, a region is defined as \emph{depleted} (or electrically active) if the local electric-field magnitude is sufficient to support efficient drift and collection of radiation-induced charge carriers under cryogenic operation. Regions exhibiting substantially reduced field magnitude are classified as weak-field or under-depleted regions, where charge transport is slower and charge collection can be incomplete for some event topologies.

\subsection{Electric field and potential distributions}

\begin{figure}
    \centering
    \includegraphics[width=1\linewidth]{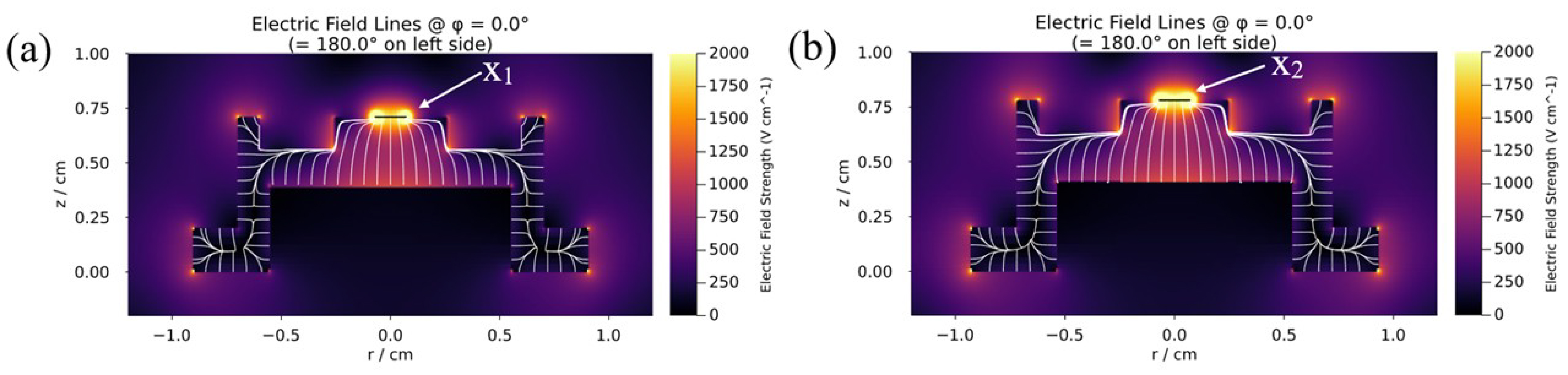}
    \caption{Simulated electric-field line distributions inside (a) SAP16 and (b) SAP17.}
    \label{Electrostatic}
\end{figure}

Figure~\ref{Electrostatic} shows the simulated electric-field line distributions for SAP16 and SAP17. In both detectors, the electric field is strongly concentrated near the point contact electrode ($X_{1}$ and $X_{2}$), where the electric potential exhibits a sharp gradient. This localized high-field region is characteristic of ICPC geometries and supports efficient signal formation at the point contact, including strong weighting-field localization for pulse-shape discrimination.

Away from the point contact, the electric field extends through the bulk toward the outer electrode, producing long carrier drift paths. Field magnitudes near the point contact approach $\sim 2 \times 10^{3}\,\mathrm{V\,cm^{-1}}$, while the bulk field is substantially lower but remains sufficient to support charge transport over most of the detector volume at the operating biases.

\begin{figure}
    \centering
    \includegraphics[width=1\linewidth]{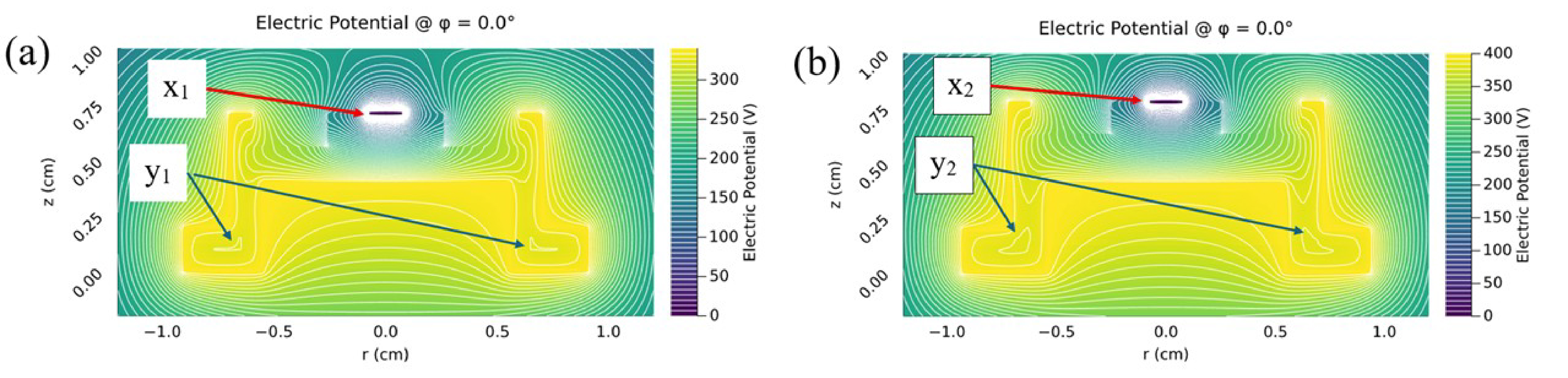}
    \caption{Simulated electric-potential field line distributions around and within (a) SAP16 and (b) SAP17.}
    \label{potential field}
\end{figure}

Figure~\ref{potential field} presents the simulated electric-potential distributions for SAP16 and SAP17. Both detectors exhibit regions of reduced potential gradient ($y_{1}$ and $y_{2}$) near the bore edge and adjacent to the wing structures, while points $x_{1}$ and $x_{2}$ indicate the point contact locations. These reduced-gradient regions correspond to localized weak-field zones in which charge carriers experience smaller drift forces and longer collection times, increasing susceptibility to trapping and ballistic deficit under finite shaping times.

\subsection{Depletion behavior and comparison with C--V measurements}

The simulated electric-potential maps provide a consistent qualitative explanation for the experimentally observed C–-V behavior discussed in Section~\ref{subsec:C-V characteristics and depletion assessment}. In both SAP16 and SAP17, regions of reduced electric field appear near the bore ledge and wing transitions, which influence the local field uniformity and the detailed evolution of the capacitance with applied bias. As discussed previously, these simulated features depend on the simplifying assumptions used in the model and therefore may not exactly reproduce the experimental behavior. 

Although SAP16 and SAP17 share the same nominal ICPC architecture, subtle geometric differences affect the spatial extent and location of these weak-field regions. In SAP17, the simulated weak-field zones extend slightly further into the detector volume at the operating bias, increasing sensitivity to charge-collection non-uniformities. This behavior is consistent with the larger intrinsic energy broadening observed in SAP17, particularly at higher gamma-ray energies (Table~\ref{tab:energy_resolution}), where local variations in charge transport contribute more strongly to peak broadening.

\begin{figure}
    \centering
    \includegraphics[width=1\linewidth]{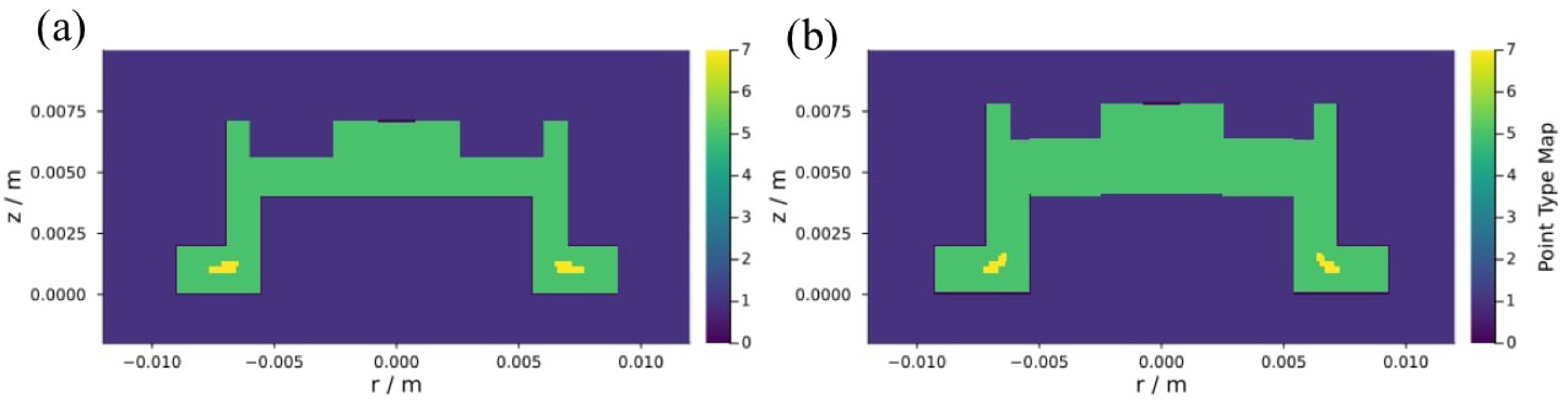}
    \caption{Simulated depletion maps of the ICPC detectors (a) SAP16 at 340 V and (b) SAP17 at 400 V, obtained using Julia-based electrostatic simulations. Green regions indicate electrically active volume, while yellow regions represent localized weak-field or under-depleted regions near the bore and wing transitions.}
    \label{active volume}
\end{figure}

The effective active fraction at the operating bias was evaluated using electrostatic simulations performed in Julia with the \texttt{SolidStateDetectors.jl} framework. The detector volume was classified based on the local electric-field magnitude, and regions exceeding a minimum field threshold required for efficient charge transport were defined as electrically active. Using this criterion, the active fraction is estimated to be approximately 96\% for SAP16 and 97\% for SAP17. The remaining fraction corresponds to localized weak-field regions near the bore ledge and wing transitions, as highlighted in Figure~\ref{active volume}.

\subsection{Polarity dependence (negative bias) and operational choice}
\label{sec:Polarity dependence (Negative bias) and operational choice}

\begin{figure}
    \centering
    \includegraphics[width=1\linewidth]{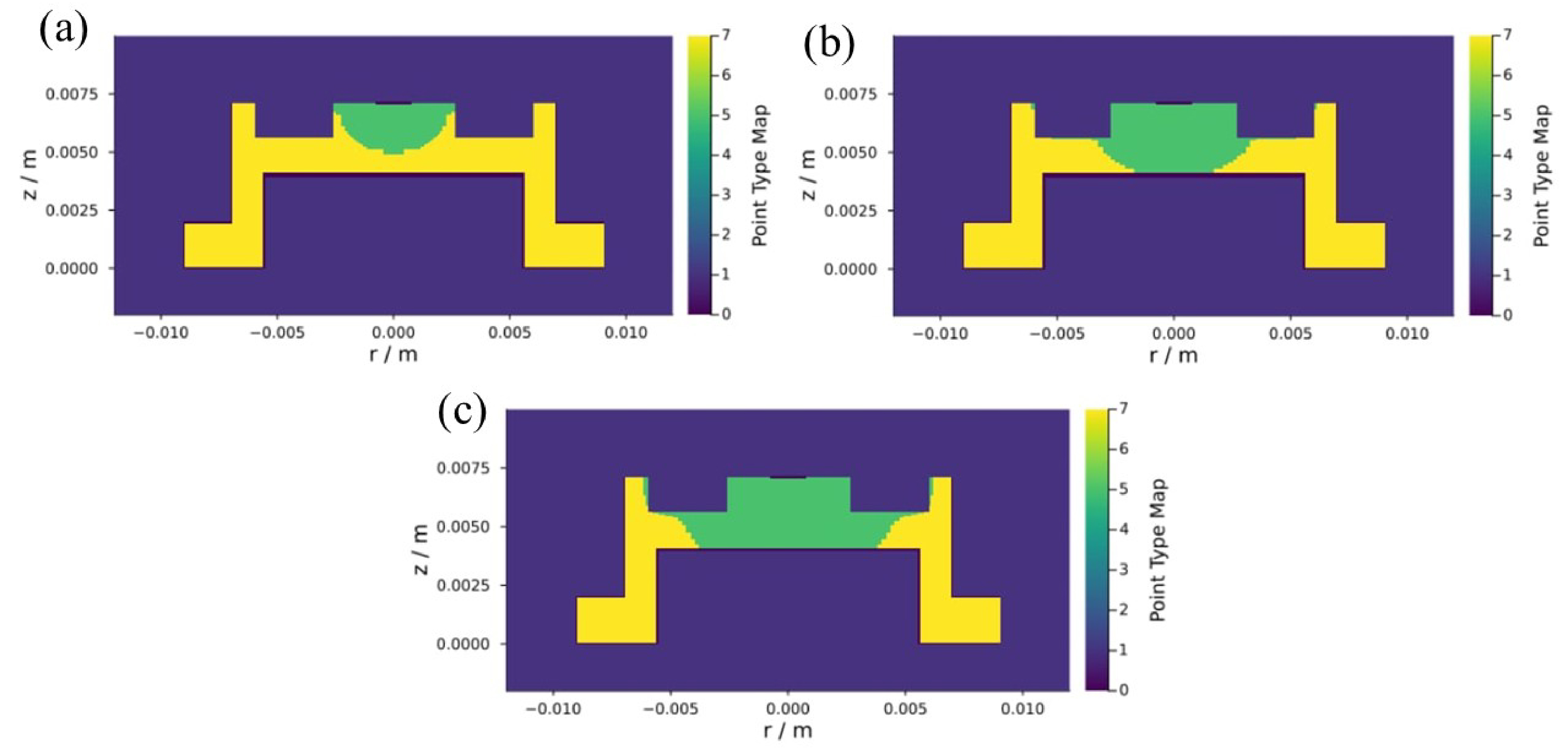}
    \caption{Simulated depletion (active-volume) maps for SAP16 under negatively biased at (a) 340 V, (b) 1000 V, and (c) 3000 V. Green regions indicate electrically active volume, while yellow regions represent localized weak-field or under-depleted regions near the bore and wing transitions.}
    \label{Negative Biased depletion}
\end{figure}

To illustrate the strong polarity dependence in the ICPC geometry, we also simulated depletion development under negative bias (i.e., $-$HV applied to the outer contact while the point contact is held at ground). In this configuration, the depleted/active region remains confined near the point contact side and does not extend efficiently into the bulk even at large reverse bias. The simulated active fraction increases only from approximately 6.67\% at 340~V to 13.84\% at 1000~V and 22.82\% at 3000~V, indicating that reversed polarity is not suitable for achieving a large depleted volume in this ICPC design. The corresponding simulated depletion maps are shown in Figure~\ref{Negative Biased depletion}.

\subsection{Simulated capacitance and comparison to experiment}

Detector capacitance was also estimated from the electrostatic simulations (Figure~\ref{Electrostatic}) by evaluating the total induced charge on the point contact electrode as a function of applied bias voltage. The simulated capacitance values at the operating biases (positive polarity) are in good agreement with experimentally extracted capacitances, as summarized in Table~\ref{tab:capacitance_comparison}. This correspondence supports the validity of the geometric implementation and boundary-condition assumptions used in the model, and it confirms that the low capacitance observed experimentally is primarily a consequence of the point contact geometry (and associated field configuration) rather than a requirement of complete volume depletion.
\subsection{Implications for detector optimization}

The combined experimental and simulation results highlight the sensitivity of ICPC detector performance to geometric features such as bore depth and diameter, wing thickness, and abrupt surface transitions. In particular, sharp geometric changes near the bore ledge and wing interfaces promote the formation of weak-field regions that can limit full depletion and degrade charge-collection uniformity, thereby increasing intrinsic peak broadening. These findings suggest that future ICPC detector designs may benefit from targeted geometric optimization, such as reducing bore diameter, smoothing or beveling bore-edge transitions, modifying wing dimensions, or introducing gradual surface contours to improve field uniformity. Systematic parametric simulations of these design variables will be essential for achieving more uniform depletion and improved spectroscopic performance in next-generation ICPC detectors.

\section{Angular response and relative efficiency}
\label{sec:Angular response and relative efficiency}

\subsection{Measurement geometry and definition of angle}
\begin{figure}
    \centering
    \includegraphics[width=1\linewidth]{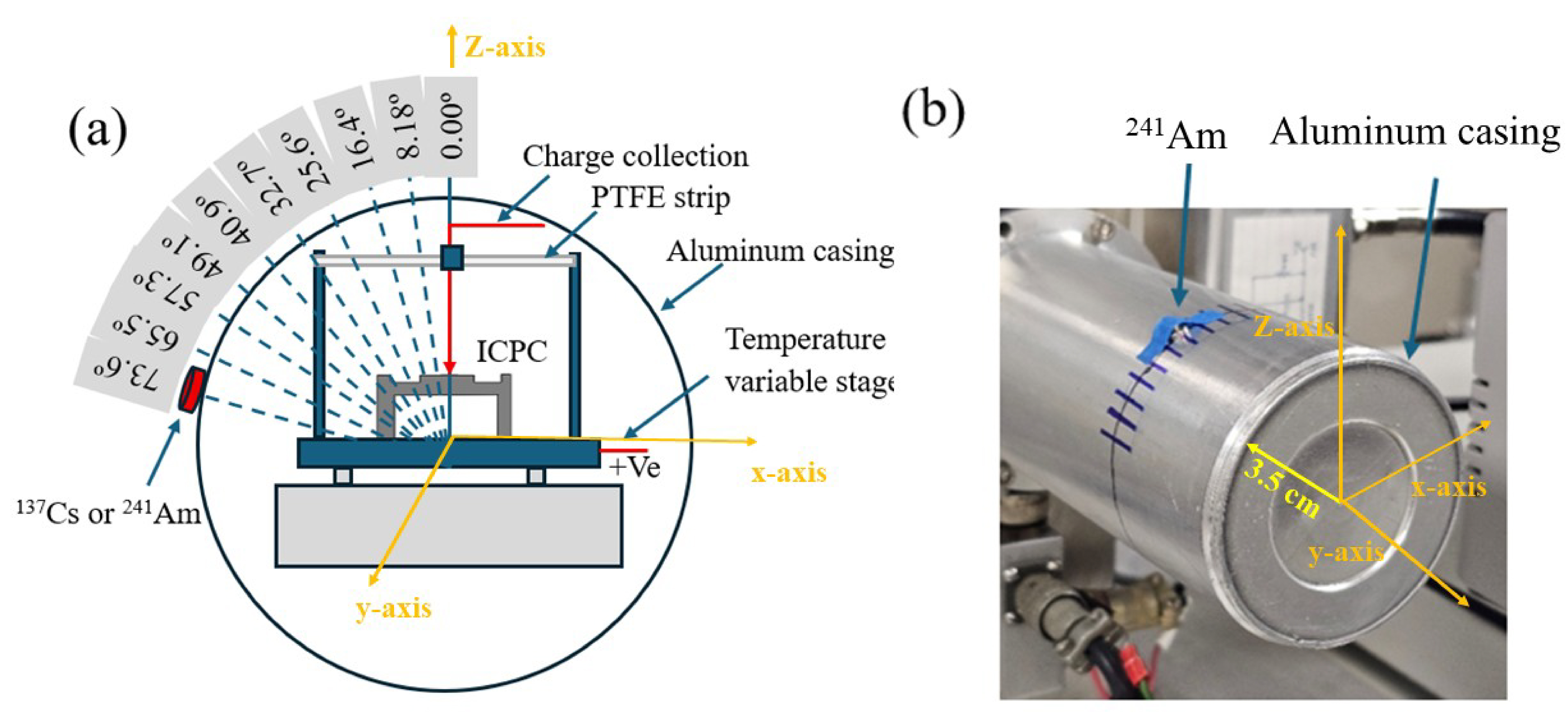}
    \caption{(a) Schematic of the angular-response measurement geometry and source rotation in the $x$--$z$ plane, and (b) photograph of the experimental setup showing the source placement relative to the detector.}
    \label{angular setup}
\end{figure}

The angular response of the ICPC detectors was investigated to quantify geometry-dependent efficiency variations, with emphasis on low-energy photons for which attenuation and near-surface effects are most pronounced. During these measurements, the detector remained fixed inside the cryostat, while the radioactive source was rotated around the detector axis within the $x$--$z$ plane at a constant source-to-detector distance of 3.5~cm. The incident angle $\theta$ is defined as the angle between the photon propagation direction and the detector axis normal to the point contact surface. With this convention, $\theta = 0^\circ$ corresponds to photons incident approximately along the bore axis, and $\theta = 90^\circ$ corresponds to photons incident perpendicular to the cylindrical sidewall. A schematic illustrating the angular definition relative to the bore, wing structures, and detector surfaces is shown in Figure~\ref{angular setup}. At each angular position, spectra were acquired under identical operating conditions, and the photopeak area was extracted using consistent fitting procedures.

\subsection{Relative efficiency extraction}
To reduce systematic uncertainties associated with absolute source activity, solid-angle normalization, and small source-position variations, the angular response is reported in terms of \emph{relative} efficiency. The relative efficiency at each angle was defined as the photopeak area normalized to the maximum observed photopeak area for the corresponding gamma-ray energy. Repeated measurements (three independent acquisitions) were performed at selected angular positions to assess reproducibility. The extracted relative efficiencies were consistent within statistical uncertainty, confirming stability of the source positioning and analysis procedure.

\subsection{Energy-dependent angular response at 59.5~keV and 662~keV}

\begin{figure}
    \centering
    \includegraphics[width=1\linewidth]{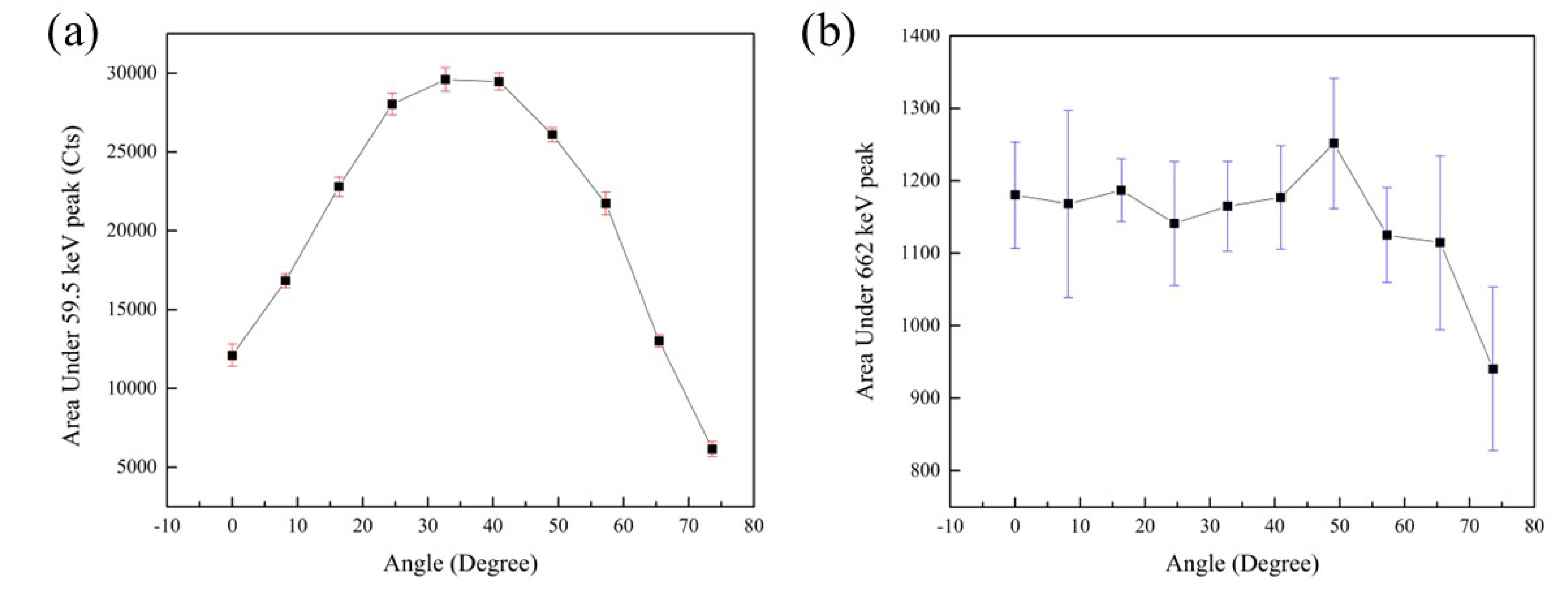}
    \caption{Relative photopeak efficiency of the ICPC detector as a function of incident angle for (a) $^{241}$Am (59.5~keV) and (b) $^{137}$Cs (662~keV). Error bars show the statistical uncertainty, taken as the standard deviation of three independent measurements at each angular position.}
    \label{angular response}
\end{figure}

Figure~\ref{angular response} shows the relative photopeak efficiency as a function of incident angle for 59.5~keV photons from a $^{241}$Am source and 662~keV photons from a $^{137}$Cs source. The angular range investigated in this study extends from $\theta = 0^\circ$ to approximately $\theta = 72^\circ$. For each energy, the photopeak area was measured at a given angle and normalized as described above, enabling direct comparison of angular trends while minimizing systematic effects related to source activity and absolute solid angle.

At 59.5~keV, photoelectric absorption dominates gamma-ray interactions in germanium \cite{Looker2014_Thesis,Berger2010_XCOM}. According to NIST XCOM data, the mass attenuation coefficient of Ge at this energy is $\mu/\rho = 2.0$--$2.3~\mathrm{cm^{2}\,g^{-1}}$, corresponding to a linear attenuation coefficient of $\mu = 11$--$12~\mathrm{cm^{-1}}$ for a Ge density of $5.32~\mathrm{g\,cm^{-3}}$ \cite{Berger2010_XCOM}. This yields an attenuation length $(1/\mu)$ of approximately 0.8--0.9~mm, implying that transmission decreases rapidly with increasing path length through Ge. Because the characteristic detector dimensions are much larger than this attenuation length, the detection efficiency becomes highly sensitive to the effective photon path length through near-surface regions and to local field/depletion conditions.

Consistent with this expectation, the 59.5~keV response exhibits a pronounced angular dependence. As summarized in Table~\ref{tab:angular_photopeak_area} in Section~\ref{sec:Appendix D}, the relative efficiency (defined from the gross counts under the photopeak) increases from $\theta = 0^\circ$, reaches a maximum at intermediate angles of approximately $\theta = 30^\circ$--$40^\circ$ (29589 $\pm$ 744 to 29467 $\pm$ 539 counts), and then decreases toward the largest measured angles (6157$\pm$ 473 counts). This behavior indicates that low-energy photons are most efficiently detected when their trajectories minimize traversal through inactive material and weak-field regions associated with the bore and geometric transitions.

Photons incident near the detector axis ($\theta \approx 0^\circ$) must traverse the bore region before reaching the bulk, and they are therefore more likely to interact or lose energy in regions affected by the bore geometry and adjacent weak-field zones, reducing the observed photopeak area. At larger angles ($\theta \gtrsim 60^\circ$), photons enter through regions that can correspond to increased effective material thickness and electric-field non-uniformity near the sidewall and wing structures, which can further suppress full-energy collection via partial charge collection or enhanced surface-related losses. At intermediate angles, the photon path length through these limiting regions is reduced, increasing the probability that interactions occur within well-depleted, higher-field regions of the detector, producing the observed maximum in relative efficiency.

In contrast, the angular response at 662~keV is comparatively flat over the measured angular range. According to NIST XCOM data, the mass attenuation coefficient of Ge at this energy is $\mu/\rho \approx 0.07~\mathrm{cm^{2}\,g^{-1}}$, corresponding to a linear attenuation coefficient of $\mu \approx 0.373~\mathrm{cm^{-1}}$ for a Ge density of $5.32~\mathrm{g\,cm^{-3}}$ \cite{Berger2010_XCOM}. This yields an attenuation length $(1/\mu)$ of approximately 2.68~cm. At this energy, the interaction length is comparable to (or larger than) the detector dimensions and interactions are dominated by Compton scattering rather than photoelectric absorption \cite{Looker2014_Thesis}. Consequently, the probability of depositing full energy in the detector and the associated charge-collection efficiency exhibit only weak dependence on incident angle within the range explored.

\subsection{Implications of angular-response measurements}
The angular-response measurements demonstrate that the detection efficiency of ICPC HPGe detectors at low gamma-ray energies is strongly influenced by detector geometry and the internal electric-field distribution. The pronounced angular dependence observed at 59.5~keV highlights the sensitivity of low-energy photons to near-surface path length, inactive material, and weak-field regions near the bore and wing transitions. In contrast, the comparatively flat response at 662~keV indicates that higher-energy gamma rays are substantially less sensitive to these effects within the angular range explored.

These observations provide experimental support for the electrostatic simulations presented in Section~\ref{sec:Electrostatic simulation and interpretation}, which identify localized weak-field regions near the bore ledge and geometric transitions as limiting factors for uniform charge collection. The coincidence of reduced low-energy efficiency at small and large incident angles with simulated weak-field locations is consistent with the interpretation that incomplete depletion and field non-uniformity shape the low-energy angular response.

From a detector-design perspective, the results underscore the importance of minimizing weak-field regions and unfavorable geometric transitions in ICPC designs, particularly for applications requiring uniform low-energy response. Geometric optimization strategies such as smoothing bore-edge transitions, modifying wing dimensions, and adjusting bore depth/diameter may reduce angular efficiency variations and improve low-energy performance. Practically, the energy-dependent angular response also implies that detector orientation can significantly affect measurements involving low-energy photons, whereas for higher energy gamma rays the response is largely orientation independent an important consideration for low-background deployments where source distributions are not isotropic.

\section{Discussion}
\label{sec:Discussion}

The comparative performance of the two p-type ICPC detectors, SAP16 and SAP17, highlights the coupled roles of leakage current, depletion behavior, electric-field uniformity, and charge-collection dynamics in ICPC detectors employing amorphous germanium (a-Ge) dual-blocking contacts. Although both devices share the same nominal ICPC architecture and contact technology, their differing electrical and spectroscopic characteristics demonstrate that ICPC performance can be highly sensitive to subtle geometric and electrostatic differences. In particular, the measurements indicate that once leakage currents are reduced to the picoampere level, spectroscopic performance is governed primarily by the internal field configuration and the degree of charge-collection uniformity rather than by leakage current alone.

A clear trade-off is observed between electrical stability and energy resolution. SAP17 exhibits consistently lower leakage current than SAP16 across the measured bias and temperature range, indicating more effective suppression of charge injection and/or surface leakage under the chosen operating polarity. However, this advantage does not translate into improved photopeak resolution. Instead, SAP16 achieves significantly better energy resolution at both 59.5~keV and 662~keV, despite its higher leakage current. This comparison shows that, within the leakage-current regime explored here, the resolution is not limited by leakage current. Because SAP17 is thicker, it can exhibit longer average drift paths for charge carriers generated far from the point contact, which may increase sensitivity to local field non-uniformities near the bore/wing transitions and thereby contribute to photopeak broadening, particularly at higher gamma-ray energies.

The observed conditioning behavior of SAP17 further suggests that contact/interface states and trapped charge distributions can influence stability during early bias cycles. One plausible interpretation is that initial bias cycling promotes field-assisted redistribution or relaxation of trapped charge within the a-Ge layer or at the a-Ge/crystalline-Ge interface, leading to a more stable effective blocking barrier after conditioning. A second, non-exclusive possibility is that local field variations near geometric features (e.g., bore edges, groove regions, or wing transitions) alter the effective injection barrier and transport pathways along the surface, producing transient behavior during early cycles. These explanations are presented as hypotheses because the results reported here do not include time-resolved measurements or direct diagnostics of trap densities, barrier evolution, or interface-state dynamics. Additional studies involving extended time-at-bias measurements, systematic temperature dependence, or barrier extraction techniques would be required to quantitatively test these mechanisms.

Angular-response measurements provide independent experimental support for the role of geometry and field non-uniformity in shaping detector response. At 59.5~keV, the pronounced angular dependence reflects the short attenuation length of low-energy photons in Ge and the associated sensitivity to entry path, near-surface material thickness, and weak-field regions near the bore and wing structures.

The coincidence of reduced low-energy efficiency at specific angles with the locations of reduced electric-field strength identified in simulation strengthens the interpretation that electric-field non-uniformity can directly impact the low-energy response by increasing sensitivity to near-surface and low-field regions. In contrast, at 662~keV the angular response is nearly isotropic, consistent with the longer attenuation length and Compton-dominated interactions at higher energies, which reduce sensitivity to localized geometric and surface-adjacent effects within the angular range explored.

Taken together, these results provide practical design lessons for next-generation ICPC HPGe detectors. While a-Ge dual-blocking contacts effectively suppress charge injection in both SAP16 and SAP17, geometric features including bore depth and diameter, wing thickness, and the sharpness of surface transitions-ultimately govern the uniformity of the electric field and the extent of weak-field regions. The superior spectroscopic performance of SAP16 suggests that modest increases in leakage current may be acceptable if they are accompanied by improved field uniformity and a reduced weak-field volume. Future detector iterations may therefore benefit from targeted geometric refinements, such as smoothing or beveling bore-edge transitions, adjusting bore depth/diameter to reduce weak-field formation near the ledge, optimizing wing dimensions to mitigate field distortions, and improving the uniformity and robustness of surface passivation. Systematic parametric simulations, coupled with controlled fabrication trials, will be essential to identify design modifications that maximize depleted volume, improve charge-collection uniformity, and deliver consistently high energy resolution. Several limitations of the present study should also be noted. The maximum applied bias voltage was constrained by leakage-current growth and breakdown considerations, limiting the accessible bias range for C–V measurements. Although both detectors are fully depleted at their stable operating biases, these constraints prevent a detailed experimental mapping of the depletion transition solely from capacitance saturation and motivate the use of electrostatic simulations to support the interpretation of the C–V behavior.

In addition, the electrostatic simulations assume a uniform impurity concentration and idealized boundary conditions, which may not fully capture local variations in crystal quality, impurity gradients, or surface and interface states that can influence the formation of weak-field regions and charge transport.
Despite these limitations, the combined electrical, spectroscopic, angular-response, and simulation results form a consistent and physically motivated picture of the performance differences between SAP16 and SAP17, and they identify concrete pathways for optimizing future ICPC detector designs.

\section{Conclusion}
\label{sec:Conclusion}
\begin{itemize}
  \item Two p-type inverted coaxial point contact (ICPC) HPGe detectors (SAP16 and SAP17) were successfully fabricated from USD-grown, zone-refined crystals using amorphous germanium (a-Ge) dual-blocking contacts, and were evaluated via electrical measurements, gamma-ray spectroscopy, angular-response studies, and electrostatic simulations.

  \item At 76~K, stable picoampere-level leakage currents were achieved (with SAP17 exhibiting the lowest leakage current), and both detectors exhibited low capacitance of $\sim 0.5$~pF. SAP16 delivered the best spectroscopic performance, with photopeak FWHM values of 1.44~keV at 59.5~keV and 2.39~keV at 662~keV.

  \item Capacitance–voltage measurements and electrostatic simulations indicate that both detectors are fully depleted at their maximum stable operating biases, while exhibiting geometry-dependent regions of reduced electric-field strength near the bore and wing transitions, as seen in the potential-field simulations. These weak-field regions can influence charge collection and contribute to intrinsic energy broadening.

  \item Angular-response measurements demonstrate pronounced, geometry-dependent efficiency variations at low energy (59.5~keV), whereas the response at 662~keV remains largely insensitive to incident angle over the measured range.

  \item Overall, the results validate a-Ge dual-blocking contacts as a viable contact technology for ICPC HPGe detectors and indicate that the most impactful future improvements will come from geometric optimization (bore dimensions and edge transitions, wing geometry) together with enhanced passivation uniformity to minimize weak-field regions, improve depletion uniformity, and strengthen spectroscopic performance.
\end{itemize}

\section{Credit authorship contribution statement}

S.~A.~Panamaldeniya: Detector fabrication, data analysis, and manuscript writing.
K.~M.~Dong: Electrostatic simulations and simulation analysis.
D.~Mei: Conceptualization, methodology, supervision, and manuscript writing/revision.

\section*{Acknowledgements}
% The authors would like to thank... (Acknowledgements text goes here.)
This work was supported in part by the U.S. National Science Foundation under Grants No. OISE-1743790, OIA 2437416,  and PHYS-2310027, and by the U.S. Department of Energy under Grants No. DE-SC0024519 and DE-SC0004768. This research was also supported by a research center funded by the State of South Dakota. Additional support was provided by the U.S. Air Force Research Laboratory under Award No. FA9550-23-1-0495, titled “Building Artificial Intelligence Research Capacity at the University of South Dakota.” We acknowledge Lawrence Berkeley National Laboratory for providing the cryostat used for detector characterization in this work.

%\appendix
\begin{appendices}
    
\section{Capacitance and calibration}
\subsection{Capacitance extraction method and uncertainty analysis}

This appendix documents the detailed procedure used to extract detector capacitance from pulser measurements and provides an uncertainty estimate, supporting the methodology summarized in Section~\ref{subsec:capacitance_extraction}. The intent is to clarify the charge-injection calibration, the relationship between pulser amplitude and injected charge, and the dominant sources of uncertainty relevant to the capacitance values reported in the main text.

\subsection{Capacitance extraction}
Detector capacitance was determined using a calibrated pulser signal injected into the test input of the charge-sensitive preamplifier. The pulser delivers a known equivalent deposited energy $E$ at the preamplifier input, corresponding to a known number of generated electron--hole pairs $N=E/w$. This produces an injected charge $Q$ at the input, which generates a voltage step $V$ at the preamplifier output that is proportional to the total effective input capacitance.

The injected charge is
\begin{equation}
Q = \frac{E}{w}\, e ,
\label{eq:pulser_charge}
\end{equation}
where $w$ is the average energy required to create an electron--hole pair in germanium and $e$ is the elementary charge. The detector capacitance $C$ is then obtained from the ratio of injected charge to measured pulse amplitude,
\begin{equation}
C = \frac{Q}{V} = \frac{N e}{V} = \frac{E}{w} \frac{e}{V},
\label{eq:capacitance_extraction}
\end{equation}
where $V$ is the measured output pulse amplitude recorded using the oscilloscope or digitizer. All measurements were performed using identical gain settings and shaping times to ensure consistency across bias points and temperatures, and to ensure that relative capacitance trends with bias reflect detector behavior rather than electronics reconfiguration.

\subsection{Calibration procedure}
The pulser amplitude was calibrated using reference settings of the signal generator and verified through repeated measurements under identical electronics conditions. The pulser energy scale was cross-checked using known gamma-ray peaks as described in Section~\ref{subsec:capacitance_extraction}. The same electronics chain and gain configuration were maintained for all capacitance measurements presented in this work to minimize systematic variation associated with gain drift or shaping-time changes.

\subsection{Uncertainty estimation}

Uncertainty in the extracted capacitance is dominated by (i) uncertainty in the pulser energy calibration and (ii) uncertainty in the measured pulse amplitude $V$ (oscilloscope resolution and electronic noise). Treating these contributions as independent, the relative uncertainty is
\begin{equation}
\left( \frac{\delta C}{C} \right)^{2}
=
\left( \frac{\delta E}{E} \right)^{2}
+
\left( \frac{\delta V}{V} \right)^{2},
\label{eq:capacitance_uncertainty}
\end{equation}
where $\delta E$ represents the uncertainty in the pulser calibration and $\delta V$ reflects the voltage measurement resolution and noise. The combined uncertainty is estimated to be on the order of a few percent, which is sufficient for the present study because it does not affect the qualitative trends in capacitance with bias voltage behavior discussed in the main text.

\section{Additional I--V and C--V measurements (repeatability)}

This appendix presents additional current-voltage (I--V) and capacitance-voltage (C--V) measurements to demonstrate repeatability and stability of the detector electrical characteristics. Repeated I--V sweeps were performed under identical temperature and bias conditions following initial bias conditioning. After the first sweep, subsequent scans showed reproducible leakage-current behavior with minimal variation, consistent with the stabilization effects discussed in Section~\ref{subsec:I-V characteristics (leakage current versus bias voltage)}. Similarly, repeated C--V measurements acquired over the same temperature and bias ranges exhibit consistent trends across multiple sweeps.

\begin{figure}
    \centering
    \includegraphics[width=1\linewidth]{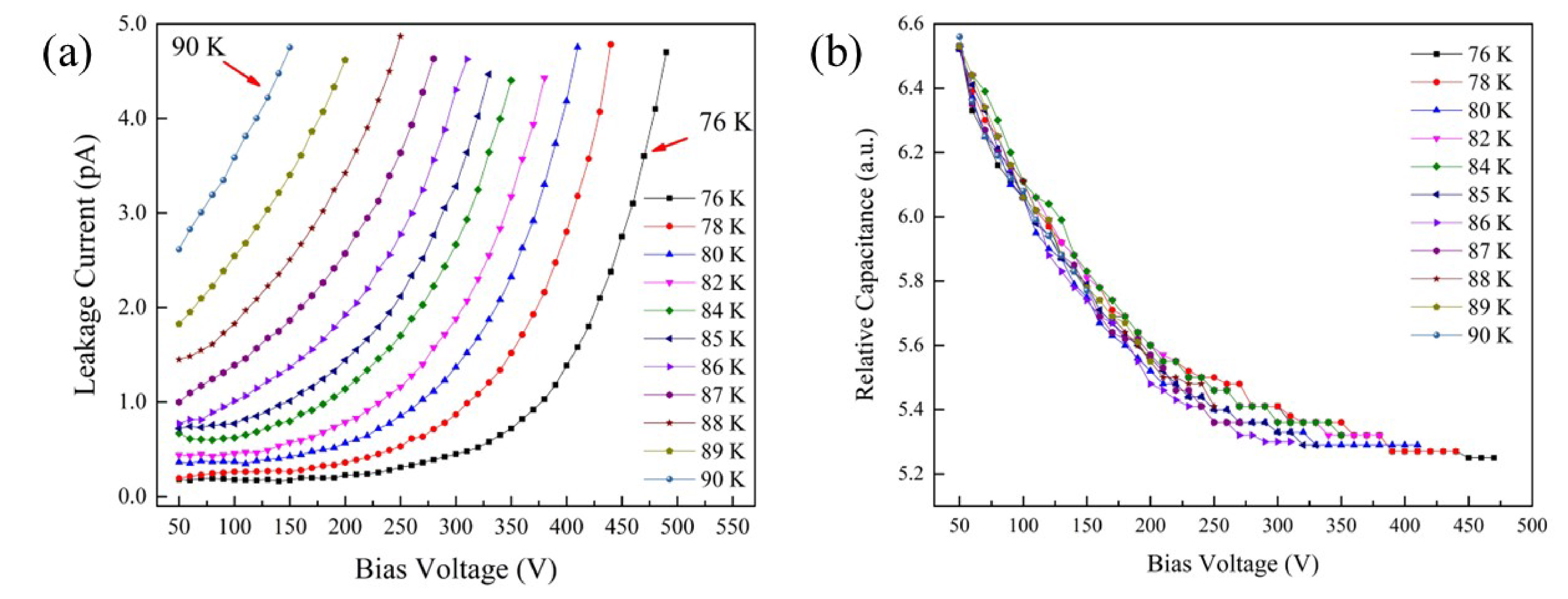}
    \caption{(a) Leakage current as a function of bias voltage measured at different temperatures ranging from 76 K to 90 K and (b) Relative capacitance as a function of bias voltage measured over the same temperature range, showing consistent depletion behavior.}
    \label{Temperature data}
\end{figure}

Figure~\ref{Temperature data} (b) shows the capacitance of the SAP17 detector as a function of bias voltage measured at temperatures between 76~K and 90~K. At all temperatures, the capacitance decreases monotonically with increasing bias, consistent with expansion of the depleted region and the associated reduction of the effective capacitance. While an abrupt plateau indicative of full depletion is not observed within the explored bias range, the overall voltage dependence is highly reproducible across temperatures. The modest temperature dependence and close overlap at higher bias voltages indicate stable electrical behavior and support the conclusions regarding leakage-current stability and depletion behavior presented in Sections~\ref{subsec:I-V characteristics (leakage current versus bias voltage)} and~\ref{subsec:C-V characteristics and depletion assessment}.

\section{Electronics chain details}

\begin{figure}
    \centering
    \includegraphics[width=1\linewidth]{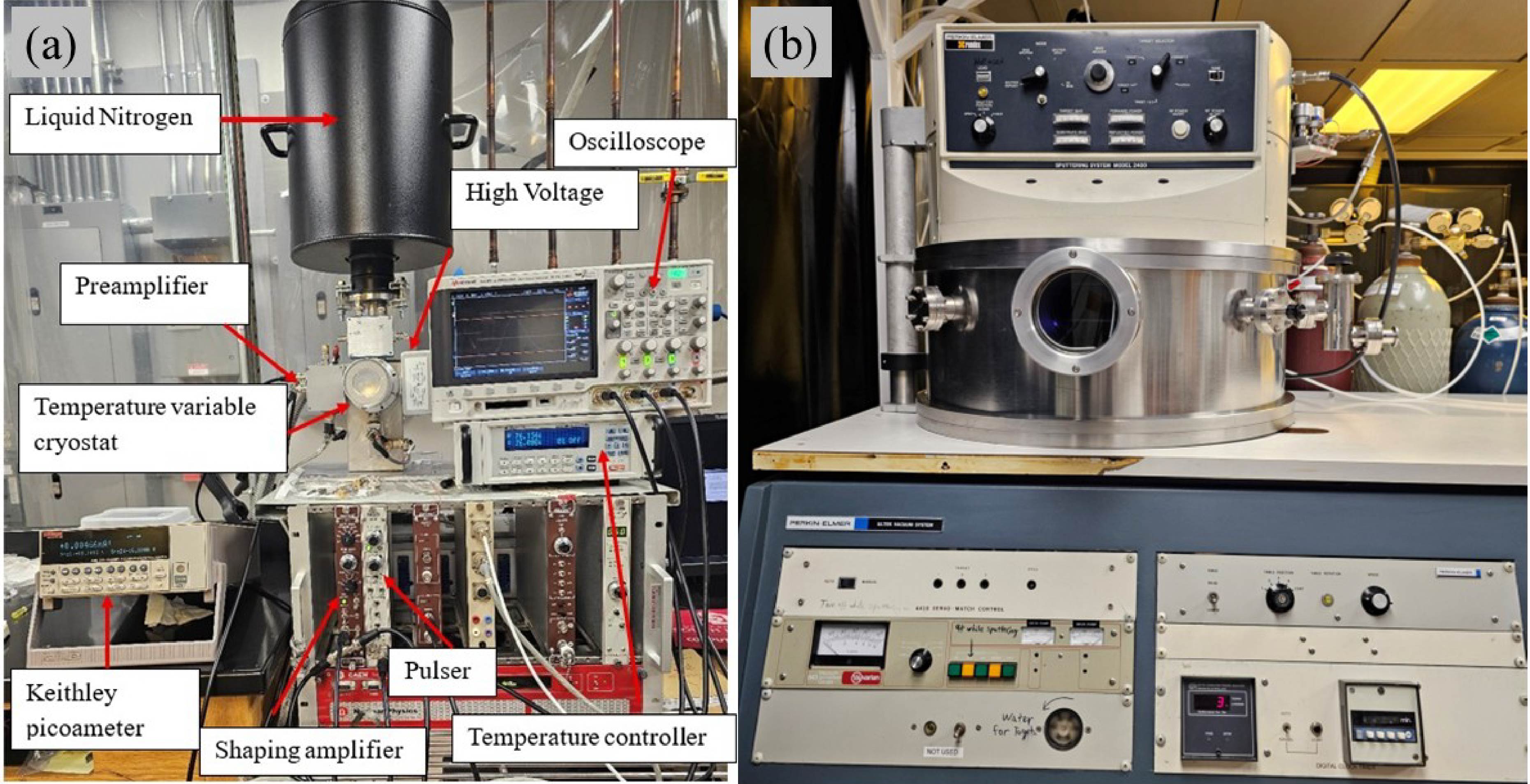}
    \caption{(a) Experimental electronics and cryogenic measurement setup used for electrical and spectroscopic characterization of the HPGe detectors and (b) PerkinElmer Model 2400 sputtering system used for amorphous-Ge and aluminum contact deposition during detector fabrication.}
    \label{Detector full setup}
\end{figure}

The detector signal was processed using a standard charge-sensitive preamplifier followed by a shaping amplifier with selectable shaping times, with a feedback capacitance $C_f = 0.6\,\mathrm{pF}$. Shaped signals were recorded using a multichannel analyzer or digitizer for spectral analysis. The same electronics configuration, gain settings, and shaping times were used consistently across all measurements to ensure comparability between detectors and operating conditions. A schematic overview of the electronics chain, including the detector, preamplifier, shaping amplifier, and data acquisition system, is shown in Figure~\ref{Detector full setup} (a) while Figure~\ref{Detector full setup} (b) shows the sputtering machine which used to make contacts.

\section{Statistical uncertainness}
\label{sec:Appendix D}

Table~\ref{tab:angular_photopeak_area} lists the measured gross photopeak areas and corresponding statistical uncertainties for the 662~keV ($^{137}$Cs) and 59.5~keV ($^{241}$Am) photopeaks as a function of incident angle. These values support the angular-response analysis discussed in Section~\ref{sec:Angular response and relative efficiency}.

\begin{table*}
\centering
\caption{Angular dependence of the gross photopeak area for $^{137}$Cs (662~keV)
and $^{241}$Am (59.5~keV) gamma rays.}
\label{tab:angular_photopeak_area}
\begin{tabular}{cccccc}
\hline
\multicolumn{3}{c}{$^{137}$Cs photopeak (662~keV)} &
\multicolumn{3}{c}{$^{241}$Am photopeak (59.5~keV)} \\
\cline{1-3}\cline{4-6}
Angle (deg) &
Gross area (cts) &
Std. dev. (cts) &
Angle (deg) &
Gross area (cts) &
Std. dev. (cts) \\
\hline
0.00  & 1180.33 & 73.36  & 0.00  & 12098.33 & 713.63 \\
8.18  & 1168.00 & 129.11 & 8.18  & 16827.00 & 452.89 \\
16.36 & 1186.67 & 43.43  & 16.36 & 22804.00 & 615.16 \\
24.55 & 1141.00 & 85.25  & 24.55 & 28035.33 & 692.50 \\
32.73 & 1164.67 & 61.78  & 32.73 & 29589.33 & 744.22 \\
40.91 & 1177.00 & 71.14  & 40.91 & 29467.33 & 539.00 \\
49.09 & 1251.33 & 89.97  & 49.09 & 26093.33 & 435.87 \\
57.27 & 1125.00 & 65.38  & 57.27 & 21727.00 & 720.94 \\
65.45 & 1114.67 & 119.82 & 65.45 & 13013.00 & 398.84 \\
73.64 & 940.33  & 112.93 & 73.64 & 6157.33  & 473.00 \\
\hline
\end{tabular}
\end{table*}

\end{appendices}

\end{document}